  \documentclass{aa}  

\usepackage{graphics}
\usepackage{graphicx}
\usepackage{multicol}
\usepackage[varg]{txfonts}

\usepackage{pdflscape}

\usepackage{natbib,twoopt}
\usepackage{amsmath} % For multiple line equations
\usepackage[breaklinks=true]{hyperref} %% to avoid \citeads line fills
\bibpunct{(}{)}{;}{a}{}{,} %% natbib format for A&A and ApJ
\makeatletter
\newcommandtwoopt{\citeads}[3][][]{\href{http://adsabs.harvard.edu/abs/#3}%
{\def\hyper@linkstart##1##2{}%
\let\hyper@linkend\@empty\citealp[#1][#2]{#3}}}
\newcommandtwoopt{\citepads}[3][][]{\href{http://adsabs.harvard.edu/abs/#3}%
{\def\hyper@linkstart##1##2{}%
\let\hyper@linkend\@empty\citep[#1][#2]{#3}}}
\newcommandtwoopt{\citetads}[3][][]{\href{http://adsabs.harvard.edu/abs/#3}%
{\def\hyper@linkstart##1##2{}%
\let\hyper@linkend\@empty\citet[#1][#2]{#3}}}
\newcommandtwoopt{\citeyearads}[3][][]%
{\href{http://adsabs.harvard.edu/abs/#3}
{\def\hyper@linkstart##1##2{}%
\let\hyper@linkend\@empty\citeyear[#1][#2]{#3}}}
\makeatother

\usepackage{color}
\definecolor{mygreen}{RGB}{0,128,0}

\hypersetup{colorlinks=true,linkcolor=blue,citecolor=blue,urlcolor=blue}
\newcommand{\muas}{$\mu$as}

\begin{document} 

\title{Gaia Data Release 1}
\subtitle{The reference frame and the optical properties of ICRF sources
%\thanks{Based on observations collected by the European Space Agency satellite Gaia}
}

\author{
F.~Mignard\inst{\ref{inst:oca}}
\and S.~Klioner\inst{\ref{inst:tud}}
\and L.~Lindegren\inst{\ref{inst:lund}}
\and U.Bastian \inst{\ref{inst:ari}}
\and A. Bombrun \inst{\ref{inst:hespaceesac}}
\and J.~Hern{\'a}ndez\inst{\ref{inst:esac}}
\and D.~Hobbs \inst{\ref{inst:lund}}
\and U.~Lammers \inst{\ref{inst:esac}}
\and D.~Michalik \inst{\ref{inst:lund}}
\and M.~Ramos-Lerate \inst{\ref{inst:vitrocisetesac}}
\and M.~Biermann\inst{\ref{inst:ari}}
\and A.~Butkevich\inst{\ref{inst:tud}}
\and G.~Comoretto\inst{\ref{inst:vegaesac}}
\and E.~Joliet\inst{\ref{inst:nasa},\ref{inst:hespaceesac}} % former developer
\and B.~Holl\inst{\ref{inst:geneva}} %ex-CU3 member; heavy contribution to AGISLab etc.
\and A.~Hutton\inst{\ref{inst:auroraesac}} %non-CU3 member with AGIS contrib 
\and P.~Parsons\inst{\ref{inst:theserverlabsesac}}
\and H.~Steidelm\"{u}ller\inst{\ref{inst:tud}}
\and A.~Andrei\inst{\ref{inst:rio}} 
\and G.~Bourda\inst{\ref{inst:bordeaux}}
\and P.~Charlot\inst{\ref{inst:bordeaux}}
}
\institute{
Laboratoire Lagrange, Universit{\'e} Nice Sophia-Antipolis, Observatoire de la C{\^o}te d’Azur, CNRS, CS 34229, F-06304 Nice Cedex, France \email{francois.mignard@oca.eu}\label{inst:oca}
\and
Lohrmann-Observatorium, Technische Universit\"{a}t Dresden, 01062 Dresden, Germany \label{inst:tud}
\and
Lund Observatory, Department of Astronomy and Theoretical Physics, Lund University, Box 43, SE-22100, Lund, Sweden  \label{inst:lund}
\and
Astronomisches Rechen-Institut, Zentrum f\"ur Astronomie der Universit\"at Heidelberg, M\"onchhofstra{\ss}e 14, D-69120 Heidelberg, Germany\label{inst:ari}
\and 
HE Space Operations BV for ESA/ESAC, Camino Bajo del Castillo s/n, 28691 Villanueva de la Ca{\~n}ada, Spain \label{inst:hespaceesac}
\and
ESA, European Space Astronomy Centre, Camino Bajo del Castillo s/n, 28691 Villanueva de la Ca{\~n}ada, Spain\label{inst:esac}
\and
Vitrociset Belgium for ESA/ESAC, Camino Bajo del Castillo s/n, 28691 Villanueva de la Ca{\~n}ada, Spain \label{inst:vitrocisetesac}
\and
Telespazio Vega UK Ltd for ESA/ESAC, Camino Bajo del Castillo s/n, 28691 Villanueva de la Ca{\~n}ada, Spain \label{inst:vegaesac}
\and
NASA/IPAC Infrared Science Archive, California Institute of Technology, Mail Code 100-22, 770 South Wilson Avenue, Pasadena, CA, 91125, USA \label{inst:nasa}
\and 
Observatoire Astronomique de l'Universit{\'e} de Gen{\`e}ve, Sauverny,  Chemin des Maillettes 51, CH-1290 Versoix, Switzerland \label{inst:geneva}
\and
Aurora Technology for ESA/ESAC, Camino Bajo del Castillo s/n, 28691 Villanueva de la Ca{\~n}ada, Spain \label{inst:auroraesac}
\and
The Server Labs S.L. for ESA/ESAC, Camino Bajo del Castillo s/n, 28691 Villanueva de la Ca{\~n}ada, Spain \label{inst:theserverlabsesac}
\and
GEA-Observatorio National/MCT,Rua Gal. Jose Cristino 77, CEP 20921-400, Rio de Janeiro, Brazil \label{inst:rio}
\and
Laboratoire d'Astrophysique de Bordeaux, Universit{\'e} de Bordeaux, CNRS, B18N, all{\'e}e Geoffroy Saint-Hilaire, 33615 Pessac, France \label{inst:bordeaux}
}

\date{Accepted on 20 September, 2016}

\abstract
  {As part of the data processing for Gaia Data Release~1 (Gaia DR1) a special astrometric
  solution was computed, the so-called auxiliary quasar solution. This gives positions for 
  selected extragalactic objects, including radio sources in the second realisation of the 
  International Celestial Reference Frame (ICRF2) that have optical counterparts bright enough 
  to be observed with Gaia. A subset of these positions was used to align the positional 
  reference frame of Gaia DR1 with the ICRF2. Although the auxiliary quasar solution was 
  important for internal validation and calibration purposes, the resulting positions are in 
  general not published in Gaia DR1.}  
  {We describe the properties of the Gaia auxiliary quasar solution for a subset of sources 
  matched to ICRF2, and compare their optical and radio positions at the sub-mas level.}  
  {Descriptive statistics are used to characterise the optical data for the ICRF sources and the 
  optical--radio differences. The most discrepant cases are examined using on-line resources
  to find possible alternative explanations than a physical optical--radio offset of the quasars.}  
  {2191 sources in the auxiliary quasar solution have good optical positions matched to 
  ICRF2 sources with high probability. Their formal standard errors are better than 
  0.76~milliarcsec (mas) for 50\% of the sources and better than 3.35~mas for 90\%. 
  Optical magnitudes are obtained in Gaia's unfiltered photometric $G$ band. The Gaia results 
  for these sources are given as a separate table in Gaia DR1. The comparison with the radio 
  positions of the defining sources shows no systematic differences larger than a few tenths of a mas. The fraction of questionable solutions, not readily accounted for by  the statistics, is less than 6\%.
  Normalised differences have extended tails requiring case-by-case investigations for around 
  100 sources, but we have not seen any difference indisputably linked to an optical--radio offset 
  in the sources. 
  }   
  {With less than a quarter of the data expected from the nominal mission it has been possible to 
  obtain positions at the sub-mas level for most of the ICRF sources having an optical counterpart 
  brighter than 20.5 mag.}

   
\titlerunning{Gaia Data Release 1 -- ICRF sources} 
\authorrunning{F.~Mignard et al.}

\maketitle

\section{Introduction} 
 
This paper presents and discusses the first Gaia astrometric solution for the optical
counterparts of radio sources in the second realisation of the International Celestial 
Reference Frame (ICRF2). This is a complementary paper to the general presentation in 
\citet{2016GaiaL} of the astrometric solutions for Gaia Data Release~1 (Gaia DR1).  
The main goal of the present paper is to provide the detection statistics for the ICRF2 subset, 
including photometric data in Gaia's unfiltered $G$ magnitude band. We also discuss and 
validate the astrometric accuracy through a straight comparison first with the defining sources 
of the ICRF2, and then including the less accurate non-defining sources. We examine a few 
individual problem sources for possible systematic optical--radio offsets. When statistically 
significant differences are found between the optical and radio positions we attempt to trace 
the root cause in the observations or data processing, such as the presence of an extended 
host galaxy or the match to a nearby star brighter than the optical counterpart, rather than 
in the sources themselves. 

The paper starts with a presentation of the data used in this investigation and how the ICRF2 
sources are matched to Gaia sources. We then discuss the overall properties of the Gaia 
solution for 2191 sources considered as optical counterparts of ICRF2 quasars. In the 
subsequent two sections we discuss the comparison of the Gaia positions to the radio ones 
respectively for the 94\% of the sources with good agreement and for the more troublesome 
cases (137 sources) where the distance between the two solutions exceeds 10~milliarcsec (mas).

\section{The data\label{data}}

\subsection{Radio positions of ICRF sources} \label{subsec:icrfsources}

The second realisation of the ICRF contains radio loud quasars observed with
very long baseline interferometry (VLBI) over a period up to 30 years (\citeads{2009ITN....35....1M}, \citeads{2015AJ....150...58F}). 
ICRF2 contains the precise position of 3414 compact radio sources and comes with an accuracy floor 
of 40~microarcsec (\muas) in each coordinate  for the whole set. ICRF2 superseded the initial 
ICRF1 solution (\citeads{1998AJ....116..516M}) with more sources and better overall accuracy. 
The two solutions are aligned to the same axes thanks to the 138 stable sources 
common to ICRF2 and ICRF1 (97 defining sources in ICRF1 and 41 other ICRF1 sources
selected for their stability).

Not all the ICRF2 sources have the same astrometric quality. The catalogue comprises three 
well delimited groups with different observation histories and statistical properties. This fact 
is very relevant for the comparison to the optical solution, since the three groups are expected
to have similar formal accuracies in the Gaia solution whereas they have very different accuracies 
in the radio positions. The division of the non-defining sources depends on whether the
source was only observed as part of the Very Long Baseline Array (VLBA) Calibrator Survey
(VCS; \citeads{2002ApJS..141...13B}) or not. The three groups of sources in 
 ICRF2 are:
\begin{itemize}
\item 
295 defining sources with a positional accuracy usually better than 0.1~mas in both 
coordinates but sometimes extending to 0.3~mas. This is generally better than the 
accuracy in Gaia DR1.
\item
922 non-VCS sources observed in several sessions, with a typical accuracy of 0.2 to 
0.3~mas but sometimes extending to several mas. About 75\% of the sources in this category have formal uncertainties smaller than in Gaia DR1.
\item 
2197 VCS-only sources observed usually in a single VCS session. These have a wide range 
of quoted accuracies from 0.2~mas to several tens of mas and a median value around 1~mas.
Within this group we may expect to find good Gaia solutions deviating from the radio positions
by $\sim$50~mas without calling for exotic physical effects in the sources.
\end{itemize}
Given the marked differences between these groups it is important to make comparisons 
per group when the accuracy of the radio positions is relevant. Particular attention should be given to the group of defining sources.

\subsection{Gaia observations of ICRF sources} \label{subsec:gaiadetections}

The optical data discussed in this paper are based on Gaia observations made
between 25 July 2014 (the start of the science mission) and 16  September 2015. During 
this period about 30~billion detections of point sources were recorded by Gaia and transmitted 
to the ground stations. During the initial processing of the data the crude celestial direction of each
detection is computed using the first approximation of the spacecraft attitude 
(see \citealt{2016GaiaF}). This step supplies the celestial position of the detection with a precision
of about 70~mas, rather independent of the magnitude (it is dominated by instrumental and 
attitude errors and not by the photon noise). This is normally sufficient to match every detection to an already known 
source, or to create a new entry in the source list. The observation of an ICRF source proceeds
in exactly the same way as for any other point-like source. As long as it is brighter than the 
detection threshold at $G \simeq 20.7$ \citep{2016GaiaP} it is likely to be detected and
observed every time it crosses the astrometric field of Gaia. During the 14~months of 
observation used for Gaia DR1 a given source was typically detected some 10 to 20 times and at each of these crossings usually accurately located on the 9 astrometric CCDs.   

Every detection is thus matched to a unique entry in the Gaia source list using the best
positional match within a window of about 1.5~arcsec as explained 
in \citet{2016GaiaF}. At the present stage of the project this matching 
is far from perfect owing to errors in the initial source list, the approximate state of the 
instrument calibrations and spacecraft attitude, and the confusion of sources in high-density
areas. Nevertheless, the matching is usually good enough that the astrometric parameters
of the source can be computed from the ensemble of detections matched to it.
   
The observations of ICRF sources are normally matched to their radio positions listed in the Initial Gaia Source List (IGSL; \citeads{2013yCat.1324....0S}) which incorporates the data of 
the Gaia Initial Quasar Catalogue (GIQC; \citeads{2014jsrs.conf...84A}) that in turn includes the ICRF2 sources with their radio coordinates. When the match is 
successful the observation is linked to the Gaia source identifier assigned to 
the ICRF source in the source list. It is important to stress that this matching is done 
during the cyclic processing, after all observations have been collected. The limitations of this cross-match are discussed in Sect.~6 of \citet{2016GaiaF}.

In this early phase of the mission the automatic photometric recognition of quasars (\citeads{2013A&A...559A..74B}) is not 
yet activated and ICRF sources are identified only from the positional match to their VLBI 
coordinates. This is not a fool-proof method since the 1.5~arcsec window used in the process 
may yield spurious identifications when the optical counterpart of the radio source is too 
faint and a field star visible to Gaia happens to be nearby and falls into the match window. ICRF sources at low galactic latitudes 
have a higher probability of being erroneously linked to Gaia observations of faint galactic 
stars. 

Out of the 3414 ICRF2 sources, a tentative optical match was obtained at least once for 
$\sim$2750 sources. But near the Gaia detection limit around $G=20.7$ a source is 
not necessarily detected in every field crossing depending on the noise, the location in the field 
of view, etc. If we consider the sources with repeated matches at different periods of time, 
the number of ICRF2 sources for which we can reasonably state that Gaia sees an optical 
counterpart is around 2500  (73\% of the ICRF2 sources), among them 270 defining sources. 
The proportion of detections decreases from the defining sources to the VCS-only, just 
because on the average the optical counterparts are slightly fainter. These numbers based 
on the first dataset fully processed should not change drastically when more data come in 
the next data release, since the sky has been scanned already three times over the 14~months 
and most detectable sources are already in the database. The sources close to the detection 
threshold will remain under-observed and the astrometric solution difficult to obtain also
in subsequent releases. The final selection of ICRF sources for the present study is
described in Sect.~\ref{sect:finalSelection}.

\subsection{The auxiliary quasar solution of Gaia data} \label{sect:auxsol}

The present optical positions of ICRF sources come, with one exception, from the 
so-called auxiliary quasar solution described in Sect.~4.2 of \citet{2016GaiaL}.%
\footnote{The exception is the Hipparcos quasar 
HIP~60936 = 3C273 = ICRF J122906.6+020308, where the position comes from the 
primary (Tycho--Gaia) astrometric solution of Gaia DR1. For simplicity we ignore
this exception in the following, and refer to the data as coming from the auxiliary
quasar solution.\label{fn:3C273}} 
The auxiliary quasar solution allowed the
positions and parallaxes to be estimated for some 135\,000 known quasars, 
utilising that their proper motions can be assumed to be negligible in the current 
context. The results of this solution were used for calibration and validation 
purposes in the processing towards Gaia DR1, and for aligning the reference frame 
of Gaia DR1 to the ICRF as described Sect.~4.3 of \citet{2016GaiaL}. Constraining the parallaxes to zero is in principle better for the quasars, as discussed in \citet{2016A&A...586A..26M}, but would have deprived us of the possibility to use this solution to check, and even calibrate, the parallax zero-point, and we considered this objective as more important than  the marginal improvement obtained by forcing the quasar parallax to be zero. Taking only the zero parallax and solving for positions and proper motions would not work as well given the short time base of the Gaia DR1.

However, apart from the 2191 positions discussed here, the astrometric
results of the auxiliary quasar solution are not published as part of Gaia DR1. 
The positions of quasars in Gaia DR1 are instead derived by means of the 
so-called secondary solution (Sect.~4.4 of \citeads{2016GaiaL}), which does not 
require the prior knowledge of the source being a quasar and is therefore
applicable to both quasars and galactic stars. Most sources in Gaia DR1 fainter 
than $V\simeq 11.5$ have positions from the secondary solution. 

For the ICRF sources there are thus two sets of positions in Gaia DR1: the ones
from the secondary solution available in the main catalogue of Gaia DR1
(see Appendix~\ref{sect:secondaryMatches}), 
and the ones obtained in the auxiliary quasar solution, which are available as a
separate table in Gaia DR1 and discussed in this paper. The differences between
the two sets of positions is usually less than 1~mas, but exceeds 100~mas for 
eight sources.

In the auxiliary quasar solution, where the proper motions are constrained to small 
values, there are effectively fewer parameters per source to fit than in the secondary 
solution. The auxiliary solution is therefore expected to be more accurate for sources
with small proper motions, such as the quasars. In the case of the ICRF sources
the superiority of the auxiliary quasar solution is easily shown by a direct comparison 
of the two solutions with the VLBI positions (Appendix~\ref{sect:secondaryMatches}). 
None of the optical positions presented here is therefore based on the secondary
solution. Constraining the proper motion is a compromise at this stage of the processing and in the further releases with more data available and full astrometric solution no such assumption will be necessary.

\subsection{Identification of ICRF sources in the Gaia auxiliary quasar solution} \label{sect:finalSelection}

This section deals with the identification and selection of the ICRF sources in the Gaia 
auxiliary quasar solution. The difference between this selection and the content of the 
Gaia DR1 secondary solution for ICRF sources is discussed in Appendix~\ref{sect:secondaryMatches}.

Given the auxiliary astrometric solution described in Sect.~\ref{sect:auxsol} it is possible 
to make a much more reliable identification of ICRF sources among the tentative 
matches discussed in Sect.~\ref{subsec:icrfsources}.   
With an astrometric accuracy better than 10~mas in both catalogues, the probability 
of a chance alignment with a star remains below $10^{-3}$ for each target even 
in the galactic plane where Gaia detects about $10^6$ sources per square degree. 
The worst case happens when there is no visible optical counterpart and a 
faint star is seen in the match window, since that object will very likely be selected and only 
much later in the processing one would realise that the match is dubious. 

The final set of ICRF sources used in the subsequent analysis consists of the sources 
from the auxiliary quasar solution that satisfy all of the following criteria:
\begin{itemize}
\item
number of matched detections $N \ge 4$;
\item
excess source noise%
\footnote{This measures the amount of noise, including calibration errors, 
that must be assumed to exist in the elementary observations, in addition to 
the photon noise, in order to account for the dispersion of residuals in the 
astrometric solution \citepads{2012A&A...538A..78L}.} 
$\epsilon_i < 20$~mas;
\item
positional uncertainty (semi-major axis of the formal error ellipse) 
$\sigma_\text{pos,max} < 100$~mas;
\item
angular separation from the ICRF2 position $\rho < 150$~mas.
\end{itemize}
The first three criteria roughly correspond to the quality criteria applied to the secondary
solution in Gaia DR1 (Eq.~12 in \citeads{2016GaiaL}), while the fourth filters out some sources
with adequate astrometric solutions but where the matching to ICRF2 is doubtful. The 150~mas
limit corresponds to a clear transition in the distribution of $\rho$ and is beyond the largest 
uncertainty in the ICRF2. These criteria result in a list of optical positions for 2191 sources,
of which 262 are defining sources in ICRF2, 640 are non-VCS, and 1289 are VCS-only.

% aux_qso_icrf2_match
Positions and other data for these 2191 sources are published as a separate table%
\footnote{The auxiliary quasar table is available in machine-readable form from
\url{http://archives.esac.esa.int/gaia/}}
in Gaia DR1. This "auxiliary quasar table" is distinct from all other data in the release, 
in particular the positions differ slightly from the secondary solutions for the same 
sources given in the main table of Gaia DR1. As shown in Appendix~\ref{sect:secondaryMatches}
the present data are both more accurate and more reliable than the secondary data.
The fields included in the auxiliary quasar table are:
\begin{itemize}
\item Gaia source identifier (64-bit integer);
\item reference epoch in Julian years;
\item right ascension $\alpha$ (degrees);
\item standard uncertainty in right ascension $\sigma_{\alpha*}=\sigma_\alpha\cos\delta$ (mas);
\item declination $\delta$ (degrees);
\item standard uncertainty in declination $\sigma_\delta$ (mas);
\item correlation coefficient $C$ between the errors in $\alpha$ and $\delta$ for each source;\footnote{Inter-source correlations are expected to be negligible, given the large angular separation of ICRF sources (several degrees) and the scanning geometry of Gaia. Empirically, we found that the positional differences between Gaia DR1 and ICRF2 have insignificant correlations, except on scales $\lesssim 10^\circ$, where they may reach 0.04.}
\item $G$ magnitude computed from the mean flux;
\item kind of prior used in the astrometric solution (3 = Tycho--Gaia astrometric solution, 
6 = auxiliary quasar solution);
\item ICRF designation of the source matched to this Gaia source (16-character string);
\item flag indicating how this source was used to fix the orientation of the reference frame 
of Gaia DR1 (0 = not used, 1 = only $\alpha$ used, 2 = only $\delta$ used, 
3 = both $\alpha$ and $\delta$ used).
\end{itemize}
The table contains 2191 entries. The reference epoch is always J2015.0 and no modelling of the Galactic acceleration has been introduced in this solution. See the discussion in Sect.~\ref{subsec:alignment}. 
The magnitude field is empty for 39 sources without calibrated photometry in Gaia DR1. 
Only one entry has prior 3 (see footnote~\ref{fn:3C273}).

\section{Properties of ICRF sources in Gaia DR1\label{sect:properties}}

In this section we present the overall properties of the Gaia DR1 results for
the ICRF sources as they stand, without any reference to their radio positions. 
The detailed comparison with the radio positions is deferred to Sect.~\ref{sect:comp}.

\begin{figure*}[]
       \resizebox{\hsize}{!}{%
       \includegraphics{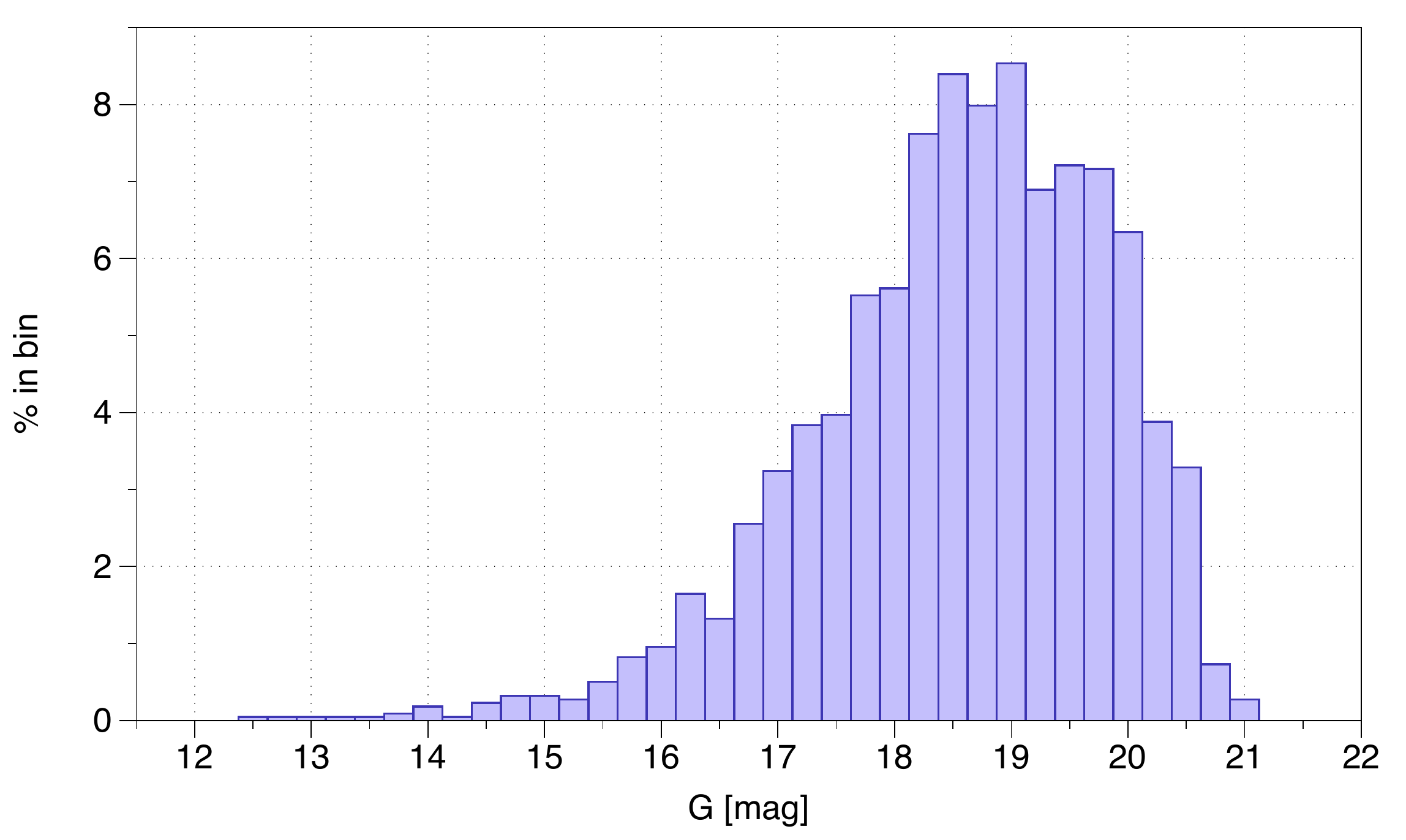}
       \includegraphics{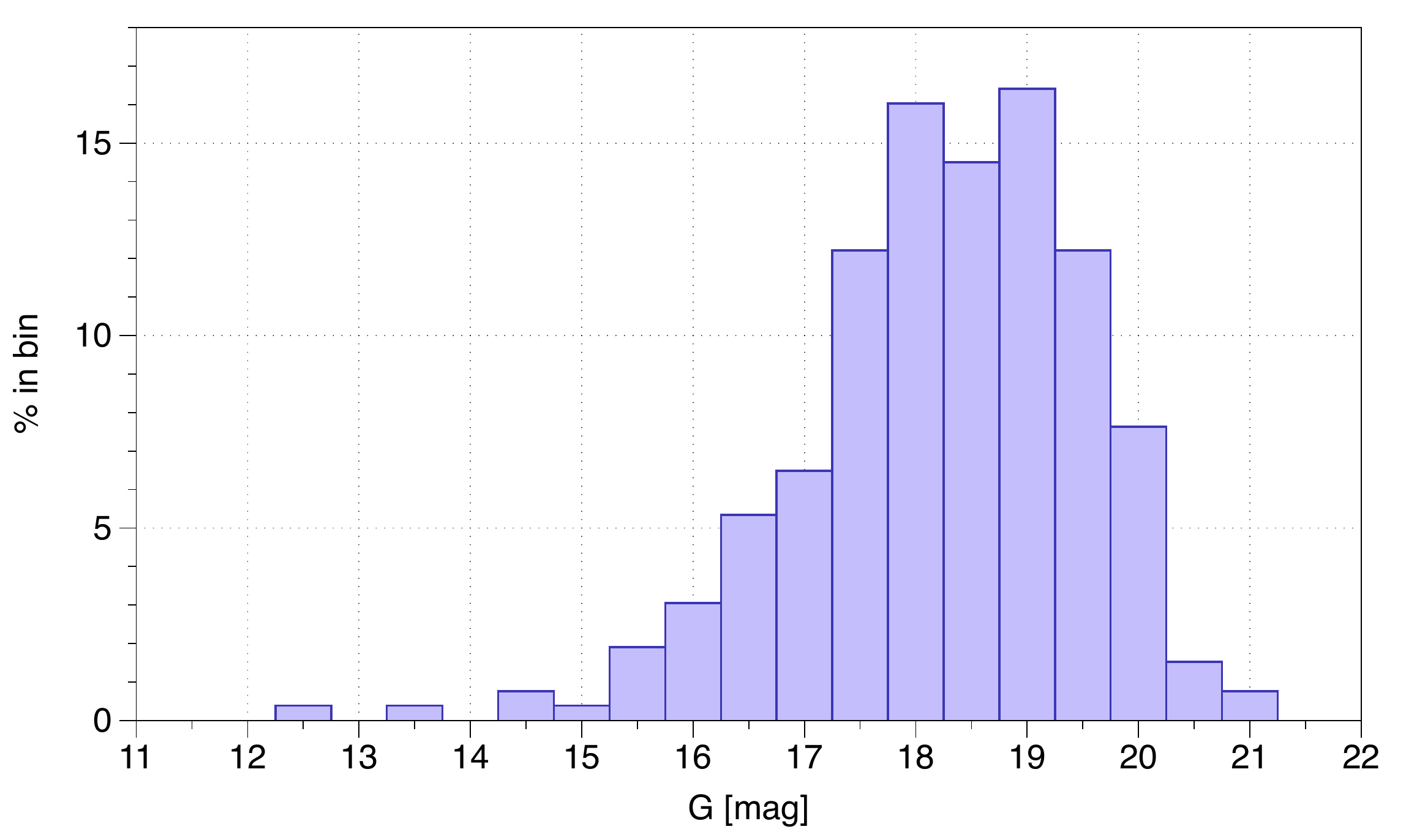}
       }
        \caption{Distribution of the $G$ magnitudes of ICRF2 sources with calibrated
        photometric data in the auxiliary quasar table.  {\it Left:} all 2152 sources. 
        {\it Right:} the 260 defining sources. \label{gmag_histo}}
\end{figure*}

\subsection{$G$ magnitudes\label{sect:magnitudes}}

An important first result derived from the data concerns the photometric properties
of ICRF sources in the Gaia $G$ band. 
For the first time, magnitudes are obtained for practically all the optical counterparts 
of the ICRF2 sources brighter than $G \simeq 20.7$. In the present list of 2191 sources there 
are 39 for which no $G$ magnitudes were derived  for Gaia DR1, although they all
have good astrometric solutions.%
\footnote{Straight inspection of the (uncalibrated) fluxes in the pre-processing for these 39 sources indicates that their 
magnitudes are in the range from 17.5 to 20.5.} 
The magnitude distribution of the remaining 2152 sources is shown in the 
left panel of Fig.~\ref{gmag_histo} and in the right panel for the subset of 260 defining 
sources. The shape of the fall-off for $G\gtrsim 20$ is at least partly an instrumental effect
due to the decreasing detection probability for fainter sources.

\begin{figure*}
       \resizebox{\hsize}{!}{%
       \includegraphics{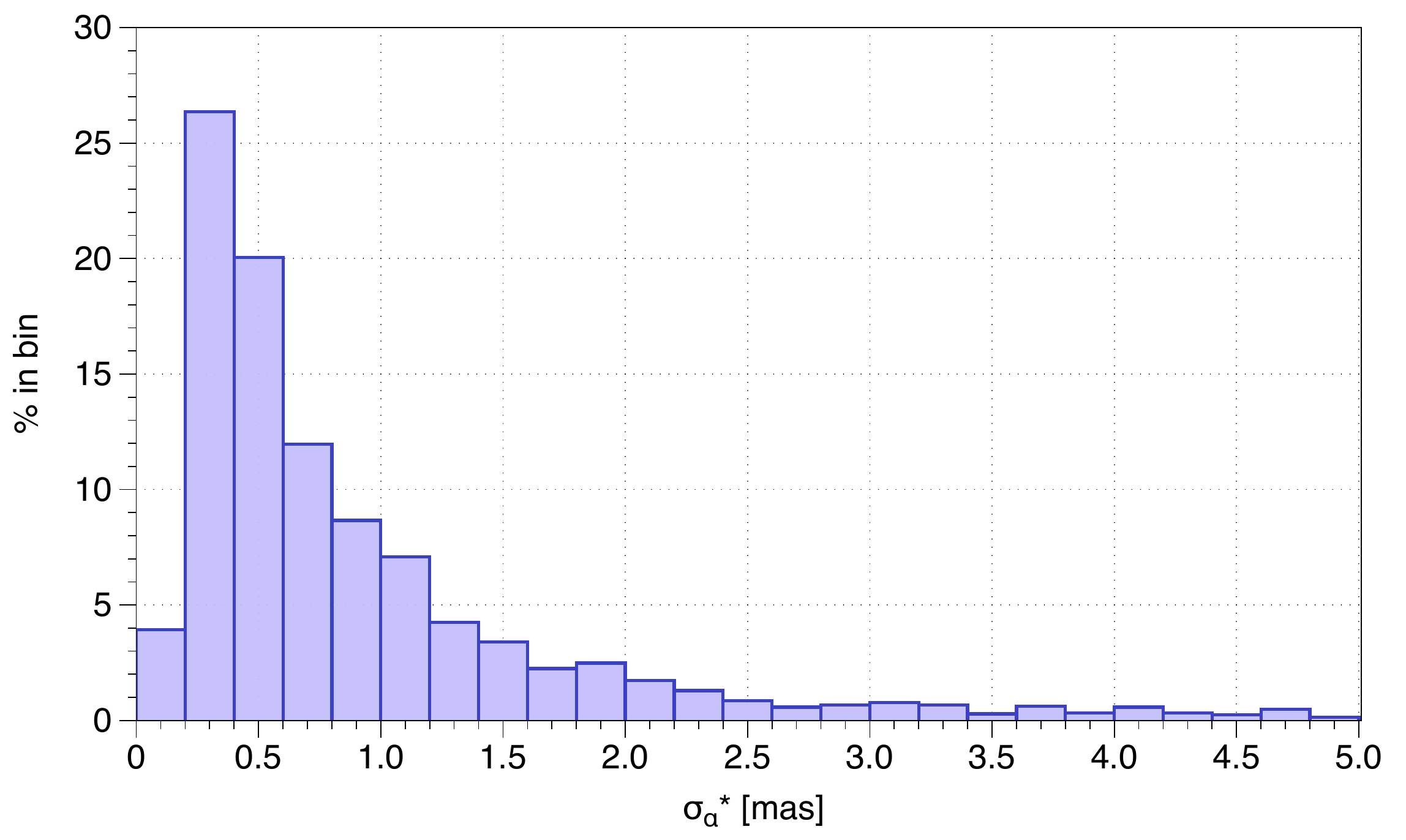}
       \includegraphics{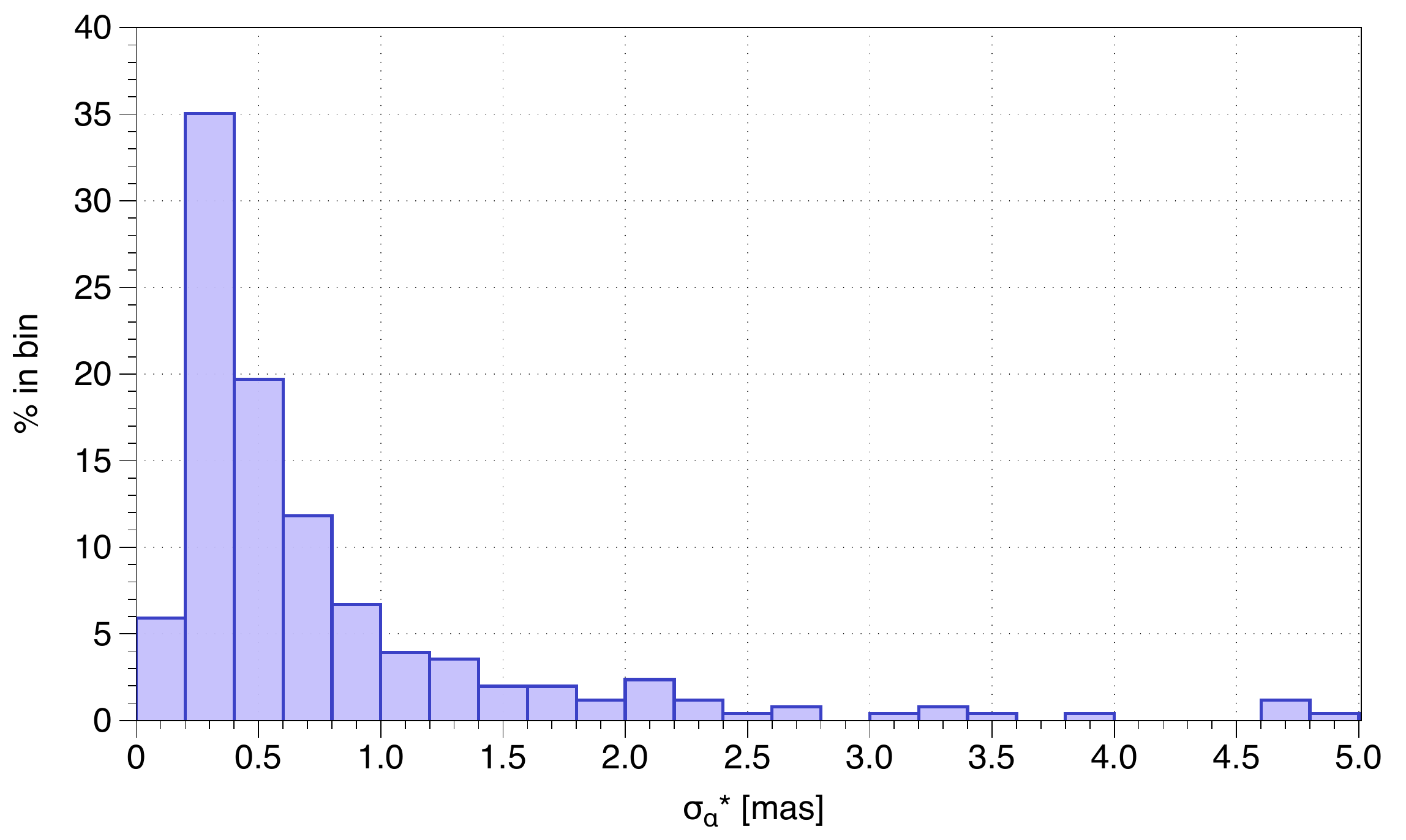}
       }
        \caption{Frequency distribution of the formal uncertainties in right ascension 
        ($\sigma_{\alpha*}$) of the optical positions of ICRF2 sources. 
        The distributions are cut at 5~mas for clarity. 
        {\it Left:} all 2191 sources (2090 with $\sigma_{\alpha*}<5$~mas).
        {\it Right:} the 262 defining sources (254 with $\sigma_{\alpha*}<5$~mas).
        \label{sra_histo}}
\end{figure*}

\begin{figure*}
       \resizebox{\hsize}{!}{%
       \includegraphics{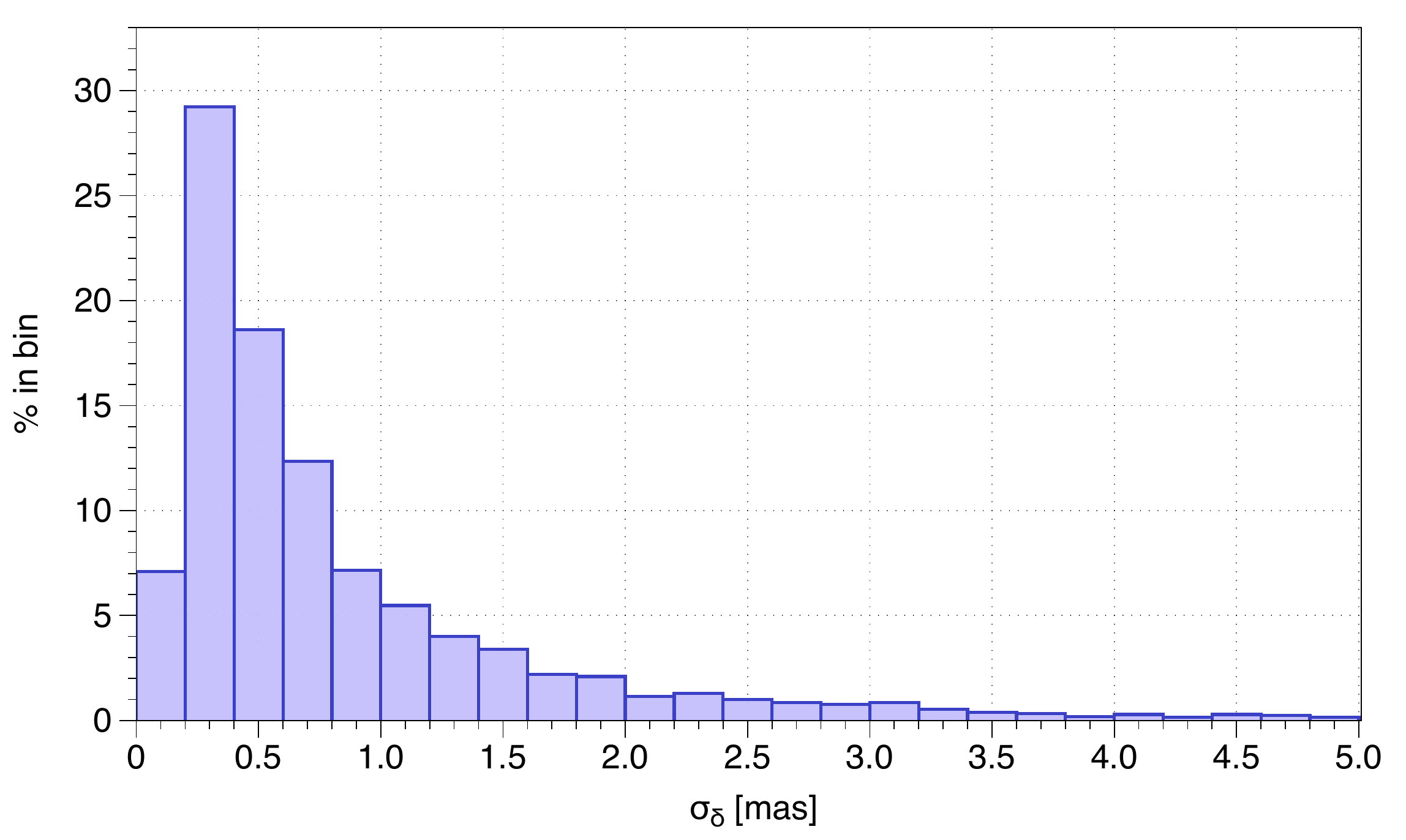}
       \includegraphics{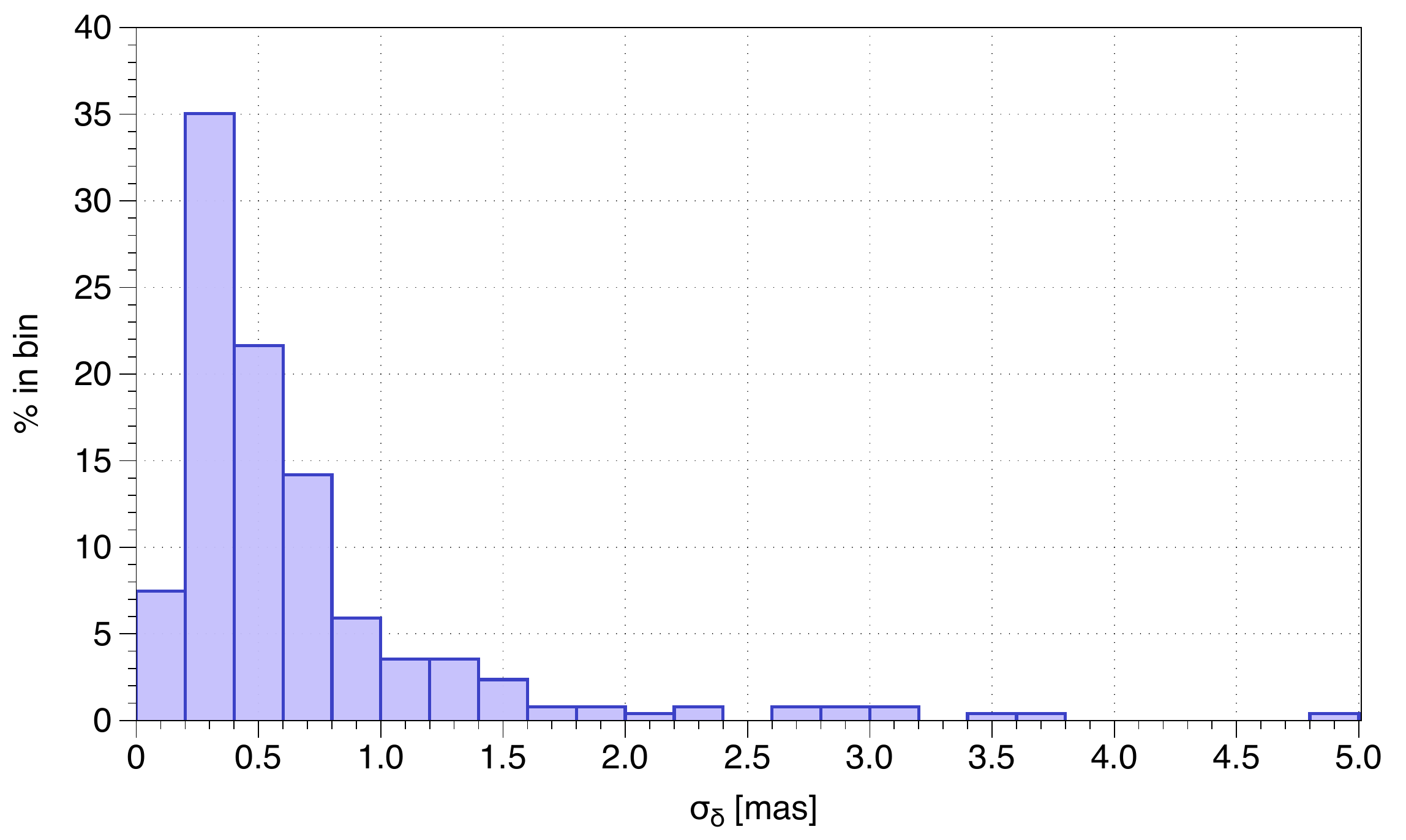}
		}
        \caption{Frequency distribution of the formal uncertainties in declination 
        ($\sigma_{\delta}$) of the optical positions of ICRF2 sources. 
        The distributions are cut at 5~mas for clarity.
        {\it Left:} all 2191 sources (2100 with $\sigma_{\delta}<5$~mas).
        {\it Right:} the 262 defining sources (254 with $\sigma_{\delta}<5$~mas).
        \label{sdel_histo}}
\end{figure*}

\subsection{Formal uncertainties of the positions\label{sect:positions}}

The distributions of the formal uncertainties $\sigma_{\alpha*}$
and $\sigma_\delta$ in the auxiliary quasar solution are shown in 
Figs.~\ref{sra_histo}--\ref{sdel_histo}, respectively.
The median value is 0.62~mas in right ascension and 0.56~mas in declination when all
2191 sources are considered, and 0.51~mas and 0.46~mas for the subset of 262 defining 
sources. These systematic differences in precision can be explained by Gaia's scanning law
and the fact that the defining sources are on average 0.3~mag brighter than the non-defining sources.

\begin{figure*}
       \resizebox{\hsize}{!}{%
       \includegraphics{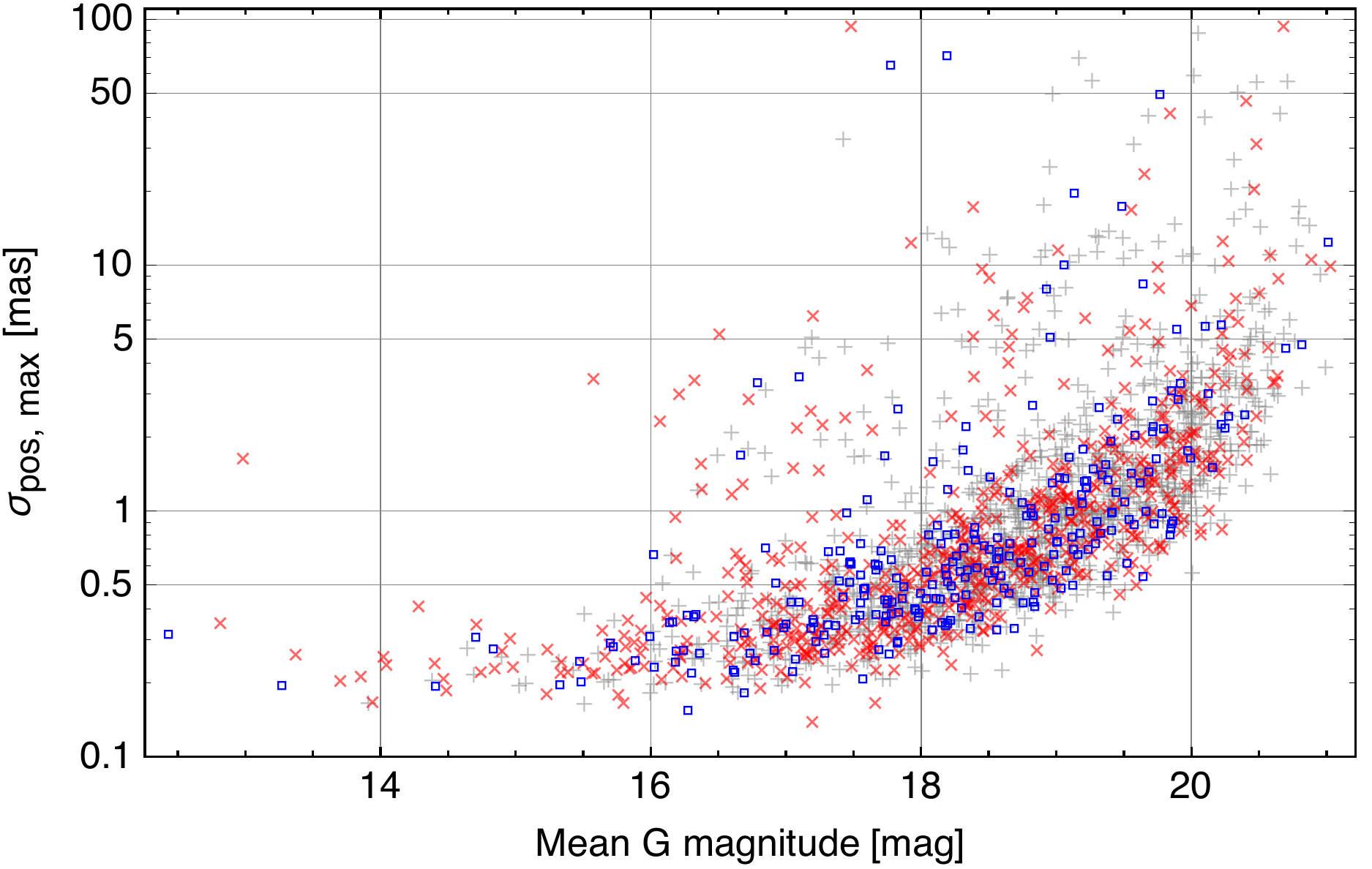}
       \includegraphics{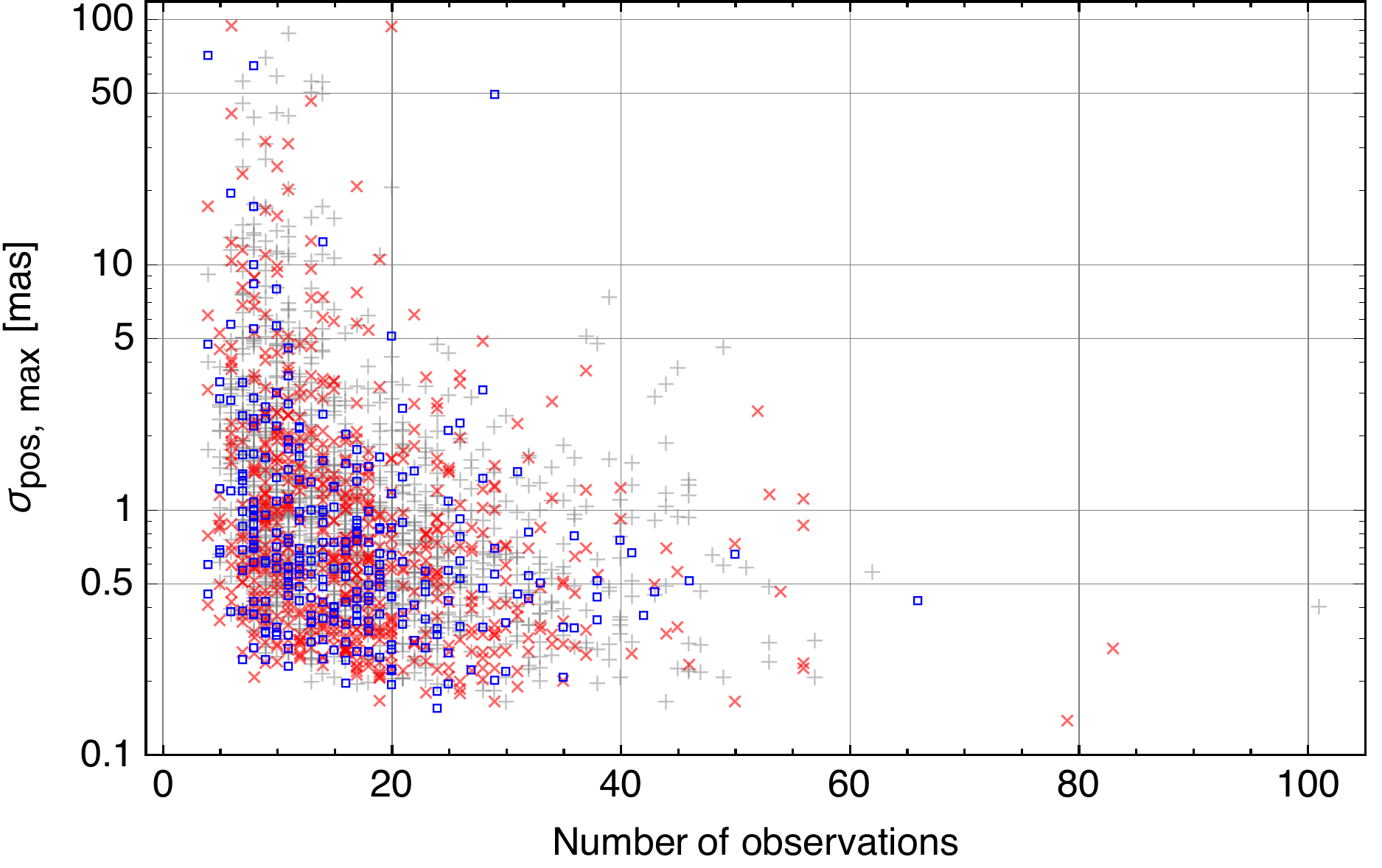}
       }
        \caption{Formal uncertainties of the optical positions of ICRF sources versus magnitude and
        number of observations per source. $\sigma_\text{pos,\,max}$ refers to the semi-major axis of the error 
        ellipse in position. {\it Left:} position uncertainty versus the $G$ magnitude for 2152 sources
        with calibrated photometry. {\it Right:} position uncertainty of the 2191 sources versus the 
        number of detections (field-of-view transits) used in the astrometric solution. 
        Symbols show defining sources as blue squares, non-VCS sources as red crosses, and
        VCS-only sources as pale grey plus signs.  
        \label{spos_gmag_nobs}}
\end{figure*}

Because of Gaia's scanning law and the relatively small number of observations used in
the present solutions, there is a high degree of correlation between $\alpha$ and $\delta$ 
for many of the sources. A better measure of the positional uncertainty could then be the
semi-major axis of the error ellipse, $\sigma_\text{pos,\,max}$, which can be computed 
from $\sigma_{\alpha*}$, $\sigma_\delta$, and the correlation coefficient between $\alpha$ 
and $\delta$ (see Eq.~9 in \citealt{2016GaiaL}).
Figure~\ref{spos_gmag_nobs} shows scatter plots of $\sigma_\text{pos,\,max}$ versus the 
$G$ magnitude (left) and the number observations (right). The three groups of ICRF sources 
are shown with different symbols.

The expected increasing trend from the photon noise is very clear for the 
fainter sources in the left panel of Fig.~\ref{spos_gmag_nobs}.
For $G<17$ there is an accuracy floor around 0.25~mas  
due mainly to the limited performance of the instrument calibration in this release. 
There remain however
some 200 sources for which $\sigma_\text{pos,\,max}$ is much higher than typical
for their magnitudes. This could have several different causes, for example that the number of 
observations, or their time distribution, is inadequate for a more accurate astrometric solution;
this in turn could be due to source confusion in the initial matching of the detections,
or to a reduced detection probability for faint sources. Source structure, including an extended 
distribution of light from the host galaxy, is another possible explanation in some cases. 
Most of the outliers in left panel of Fig.~\ref{spos_gmag_nobs} can be explained by a small 
number of observations: as shown in the right panel, large positional uncertainties are much 
more common for sources with less than 10--15 matched detections.  

\section{Comparison of the optical and radio positions}\label{sect:comp}

In this section we compare the optical positions of the ICRF sources, as obtained in the Gaia 
auxiliary quasar solution, with their reference values in the radio domain taken from ICRF2 
\citepads{2015AJ....150...58F}. Coordinate differences in right ascension and declination are 
computed as 
\begin{equation}\label{eq:dadd}
\Delta\alpha*=(\alpha_\text{DR1}-\alpha_\text{ICRF2})\cos\delta \, , \quad
\Delta\delta=\delta_\text{DR1}-\delta_\text{ICRF2} \, ,
\end{equation}
from which the angular separation between the two positions is obtained as
\begin{equation}\label{eq:rho}
\rho=\sqrt{\Delta\alpha{*}^2+\Delta\delta^2} \, . 
\end{equation}
The small-angle approximation implicit in these formulae is perfectly adequate 
for our purpose, where position differences are $<150$~mas and consequently 
second-order effects $<0.1~\mu$as.

\subsection{Position differences as angles\label{sect:angulardiff}}

We consider first the position differences expressed as angles (in mas), that is without
any scaling by their uncertainties or selection based on the uncertainties. The results are 
given in a series of diagrams showing the distributions of the coordinate differences
and positional separations both for the whole set of 2191 sources and for the subset of
defining sources. This separate consideration is mandatory given the different range of uncertainties in the ICRF2 positions for the defining or non-defining sources (Sect.~\ref{subsec:icrfsources}).

The comparison is made under the assumption that the Gaia identification actually points 
to the matched ICRF source and therefore that when a difference is found between the two positions (radio and optical) it refers to the same source. It is however clear that for the largest separations, above a few tens of mas, 
a misidentification is possible, not necessarily due to a flaw in the Gaia cross-matching, but 
more likely because of the lack of a true optical counterpart within the detection range of Gaia. 
This will be investigated in later releases when more observations have been collected. Some 
sources have a relatively large uncertainty in ICRF2 (more than 5--10~mas), which could also 
be an explanation, in particular when the optical position is more accurate. The uncertainties 
are taken into account in the normalised comparisons (Sect.~\ref{subsec:norm-compar}).

\subsubsection{Coordinate differences}\label{sect:coordDiffUnscaled}

\begin{figure*}
       \resizebox{\hsize}{!}{%
       \includegraphics[width=0.4\hsize]{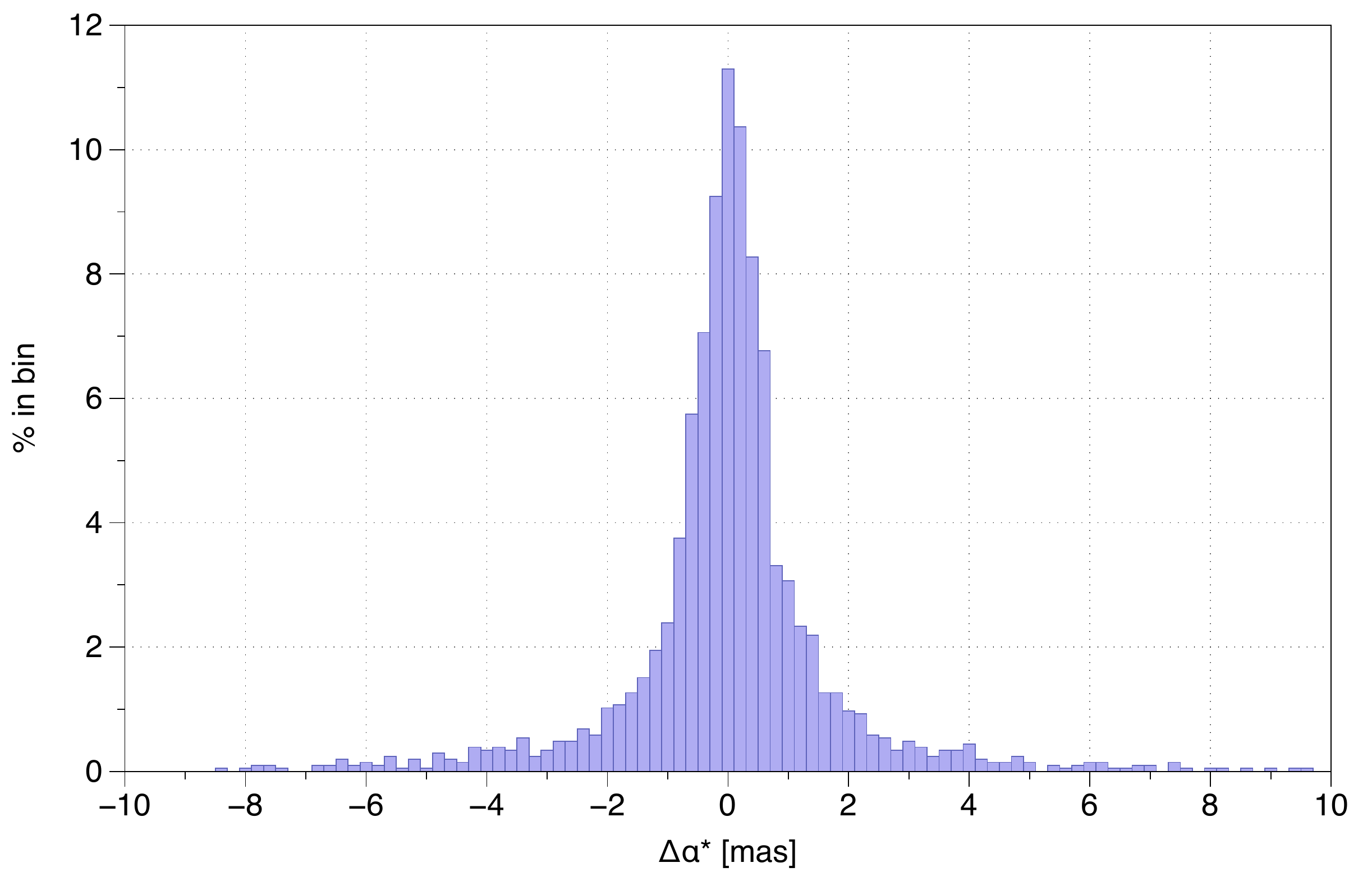}
       \includegraphics[width=0.4\hsize]{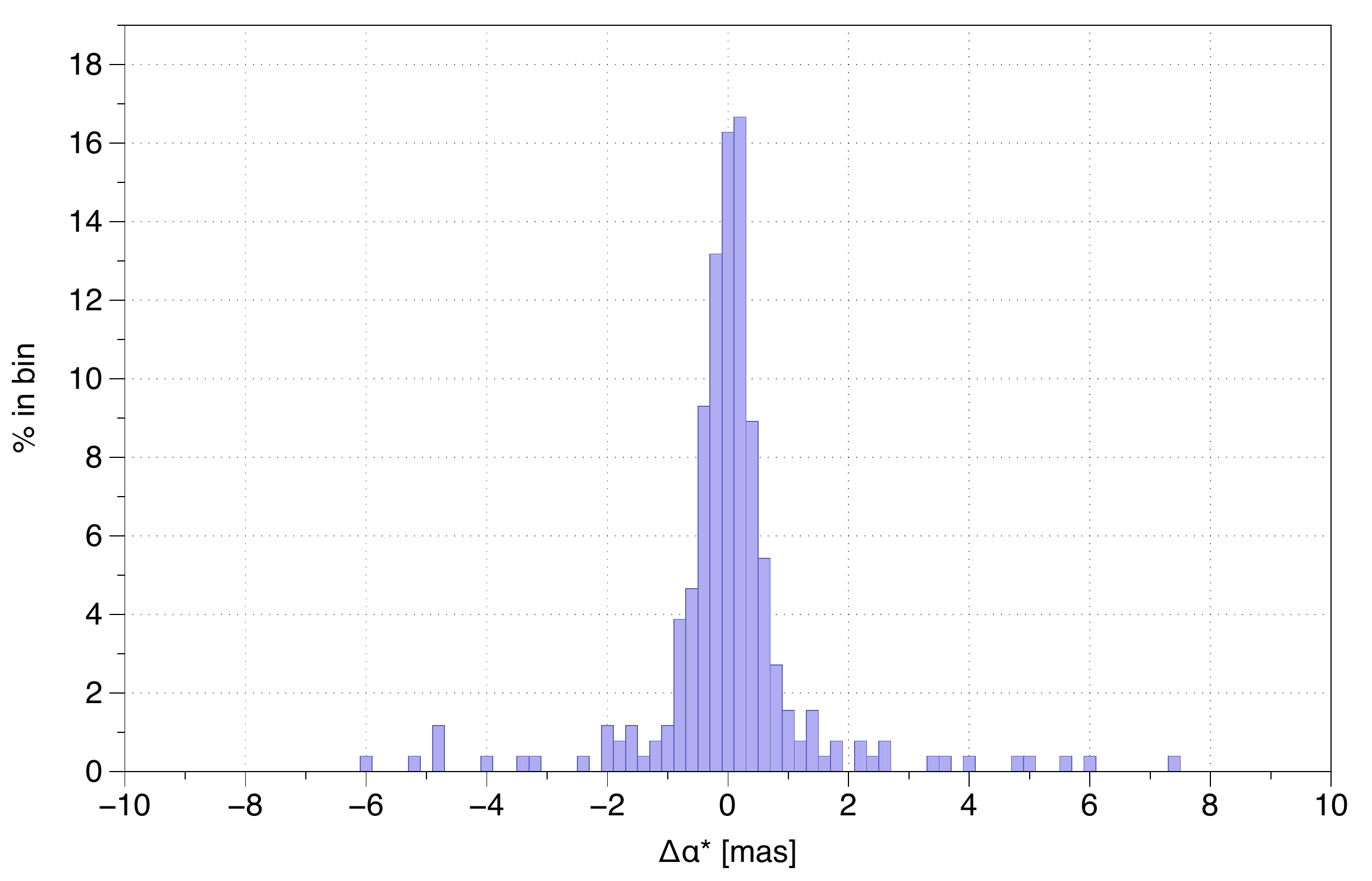}
       }
         \caption{Distribution  of the differences in right ascension between the Gaia solution and the ICRF2 positions.  {\it Left:} the 2054 solutions closer than 10~mas to their radio position.  {\it Right:} same for the 258 defining sources.
        \label{dra_10mas_histo}}
\end{figure*}

\begin{figure*}
       \resizebox{\hsize}{!}{%
       \includegraphics[width=0.4\hsize]{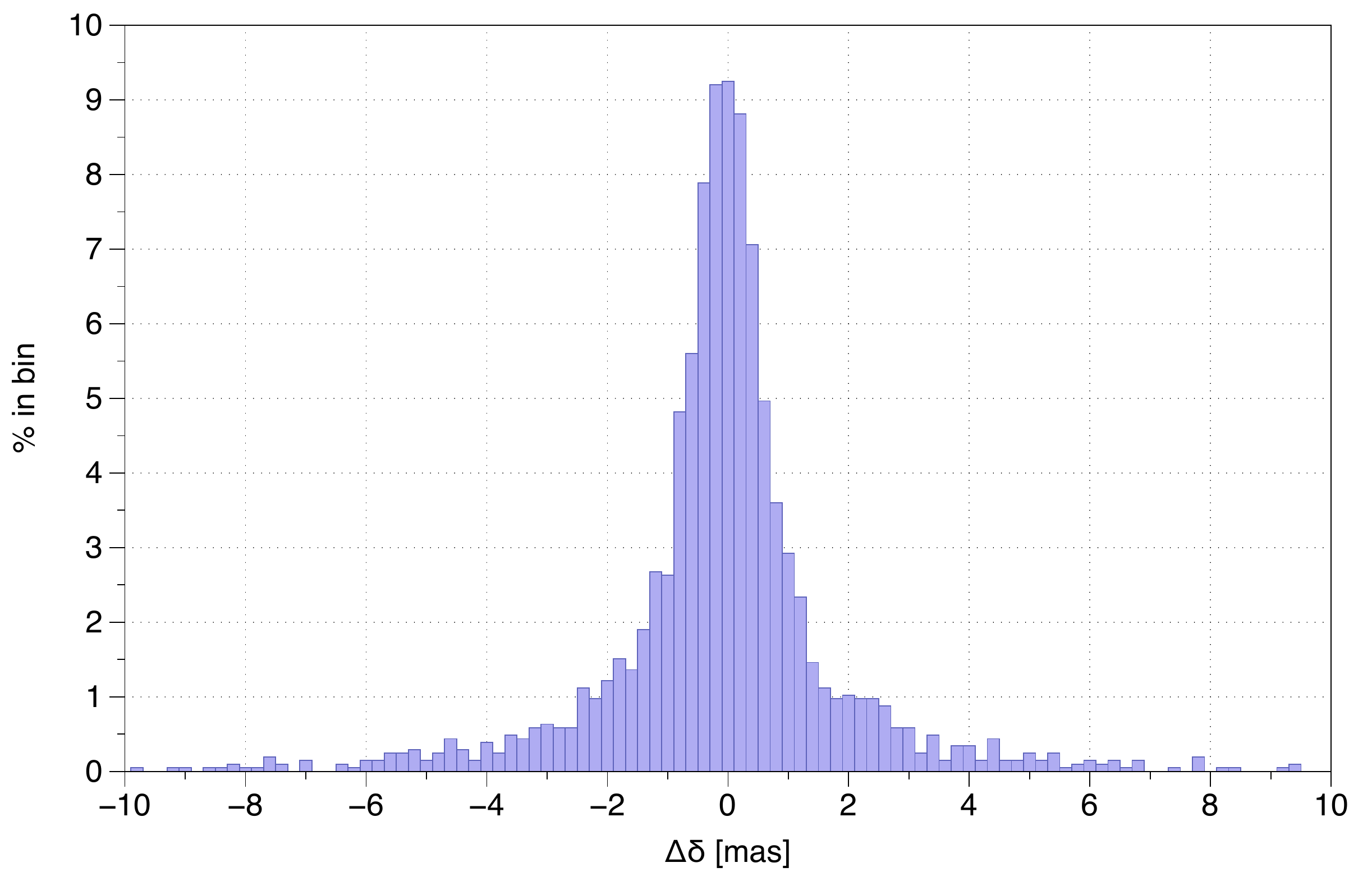}
        \includegraphics[width=0.4\hsize]{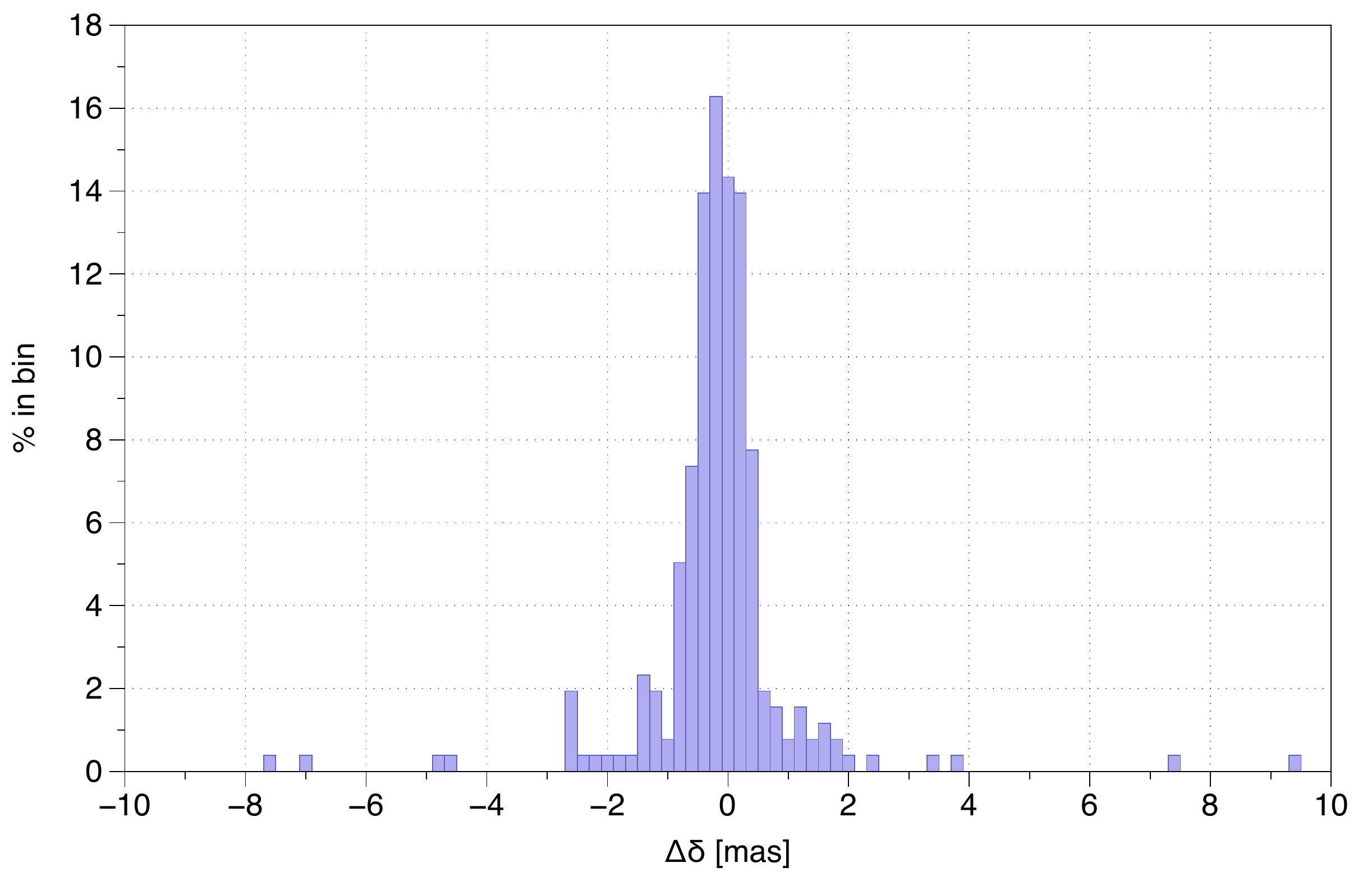}
        }
        \caption{ Distribution  of the differences in declination between the Gaia solution and the ICRF2 positions.  {\it Left:} the 2054 solutions closer than 10 mas to their radio position.  {\it Right:}  same for the  258 defining sources.
        \label{ddec_10mas_histo}}
\end{figure*}

Histograms of the differences in right ascension and declination are shown, respectively, 
in Fig.~\ref{dra_10mas_histo} and Fig.~\ref{ddec_10mas_histo}. In each figure the left 
panel shows the distribution for all sources, the right panel for the subset of defining sources.
For better visibility only the central parts (within $\pm 10$~mas) of the distributions are 
shown; the number of sources beyond this limit is given in the figure legends.

The median $\Delta\alpha{*}$ for the whole sample is $+0.038\pm 0.022$~mas. The standard 
width of the distribution, estimated by the robust scatter estimate%
\footnote{The RSE is defined as ${\left(2\sqrt{2}\,{\rm erf}^{-1}\bigl(4/5\bigr)\right)}^{-1}\approx0.390152$ times the distance between the 10th and 90th
percentiles. For a normal distribution it equals the standard deviation.} 
(RSE) is 1.50~mas, while the non-robust sample standard deviation including
the wings goes to 6.35~mas. The corresponding values for the defining subset are 
$+0.015\pm 0.038$~mas for the median, 0.74~mas for the RSE, and 3.23~mas for the sample
standard deviation. The agreement with the defining sources is obviously better owing to the 
better quality of this subset in the ICRF2. One must bear in mind that the Gaia frame has been 
aligned with the ICRF2 and that any bias in right ascension would therefore be absorbed by the 
orientation about the $z$ axis. However, the (robust) standard widths are good indicators of 
the quality of agreement between the two catalogues. 

\begin{figure*}
       \resizebox{\hsize}{!}{%
       \includegraphics[height=0.4\hsize]{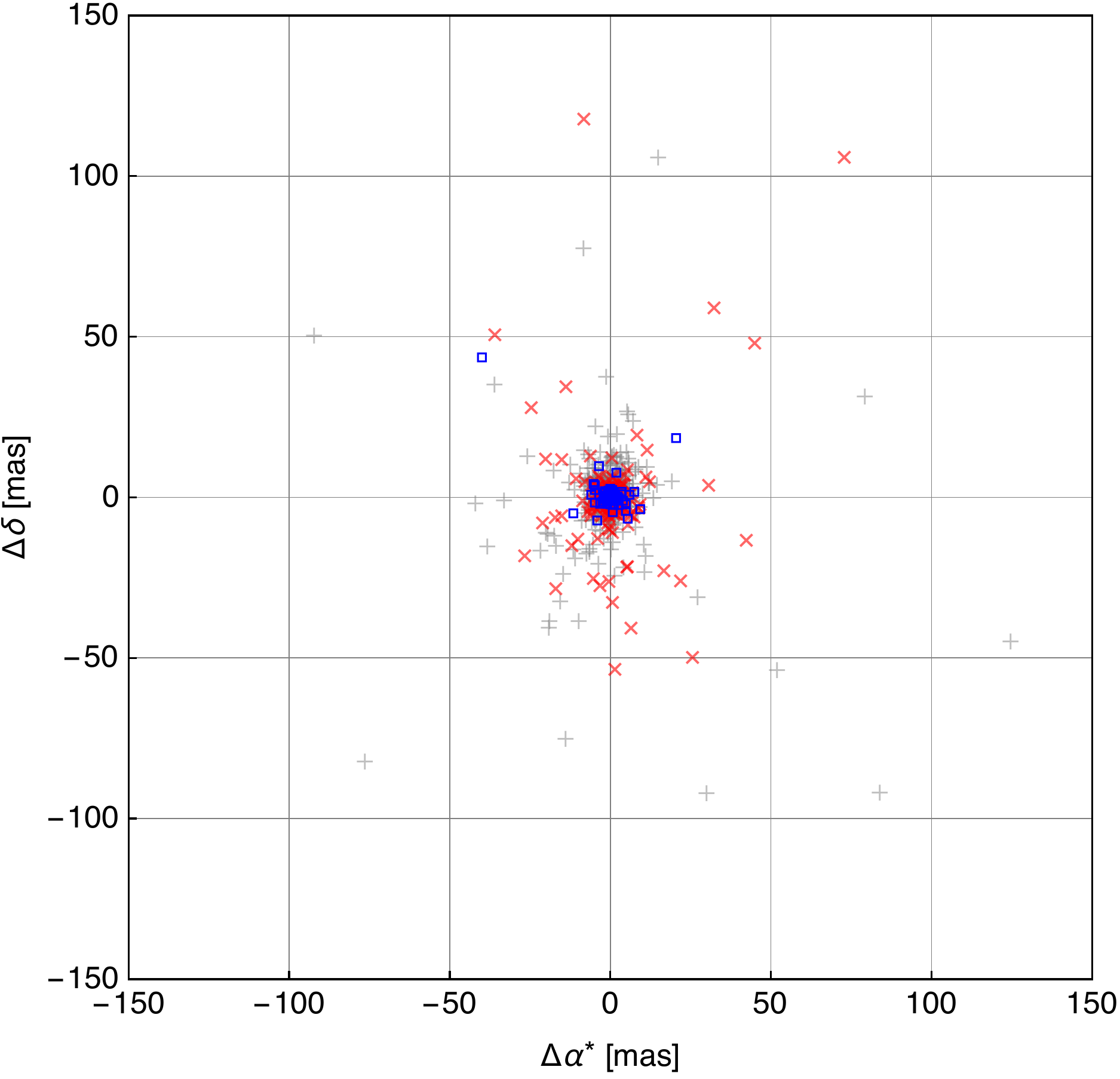}
       \includegraphics[height=0.4\hsize]{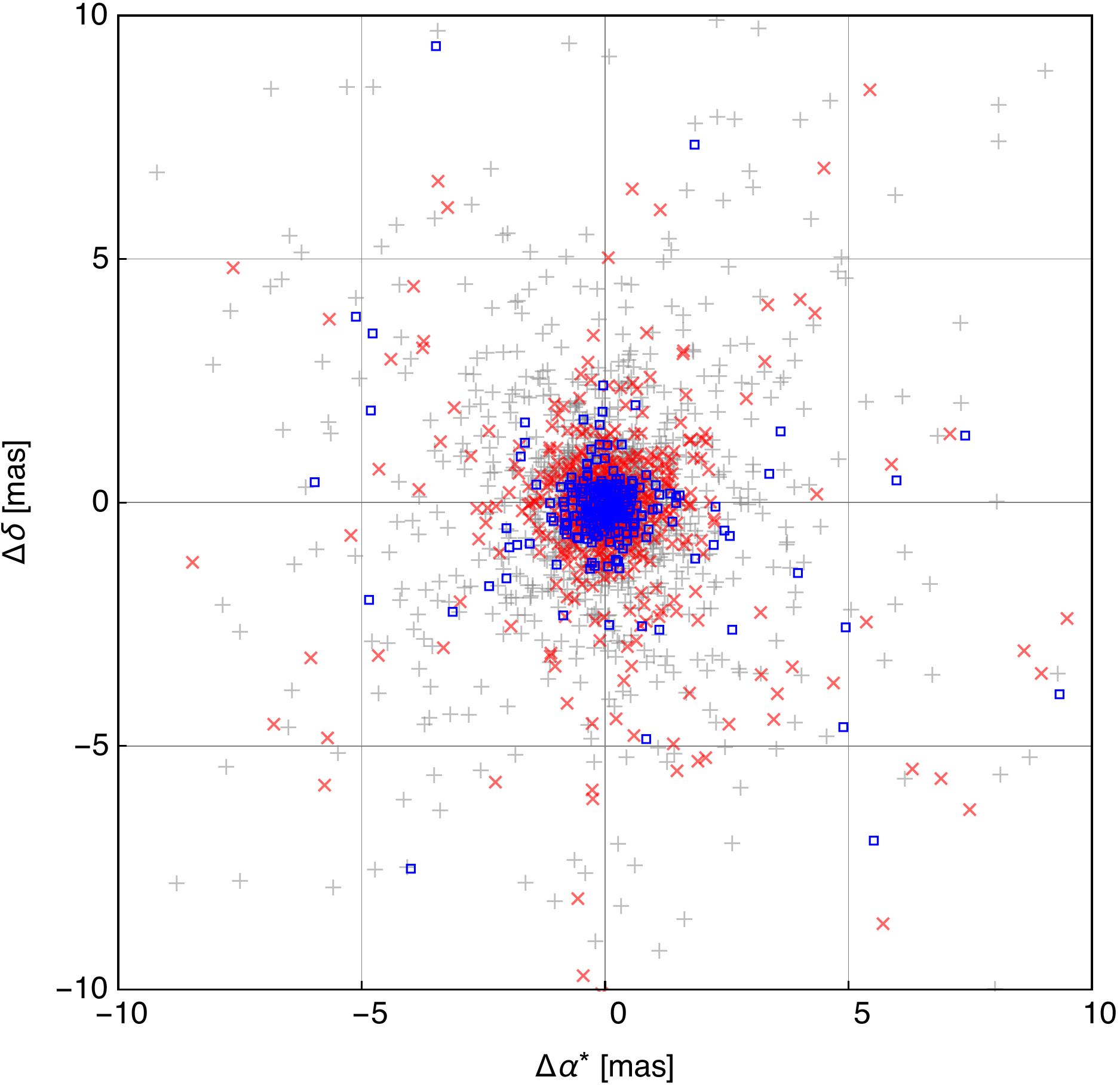}
       }
       \resizebox{\hsize}{!} {%
       \includegraphics[height=0.4\hsize]{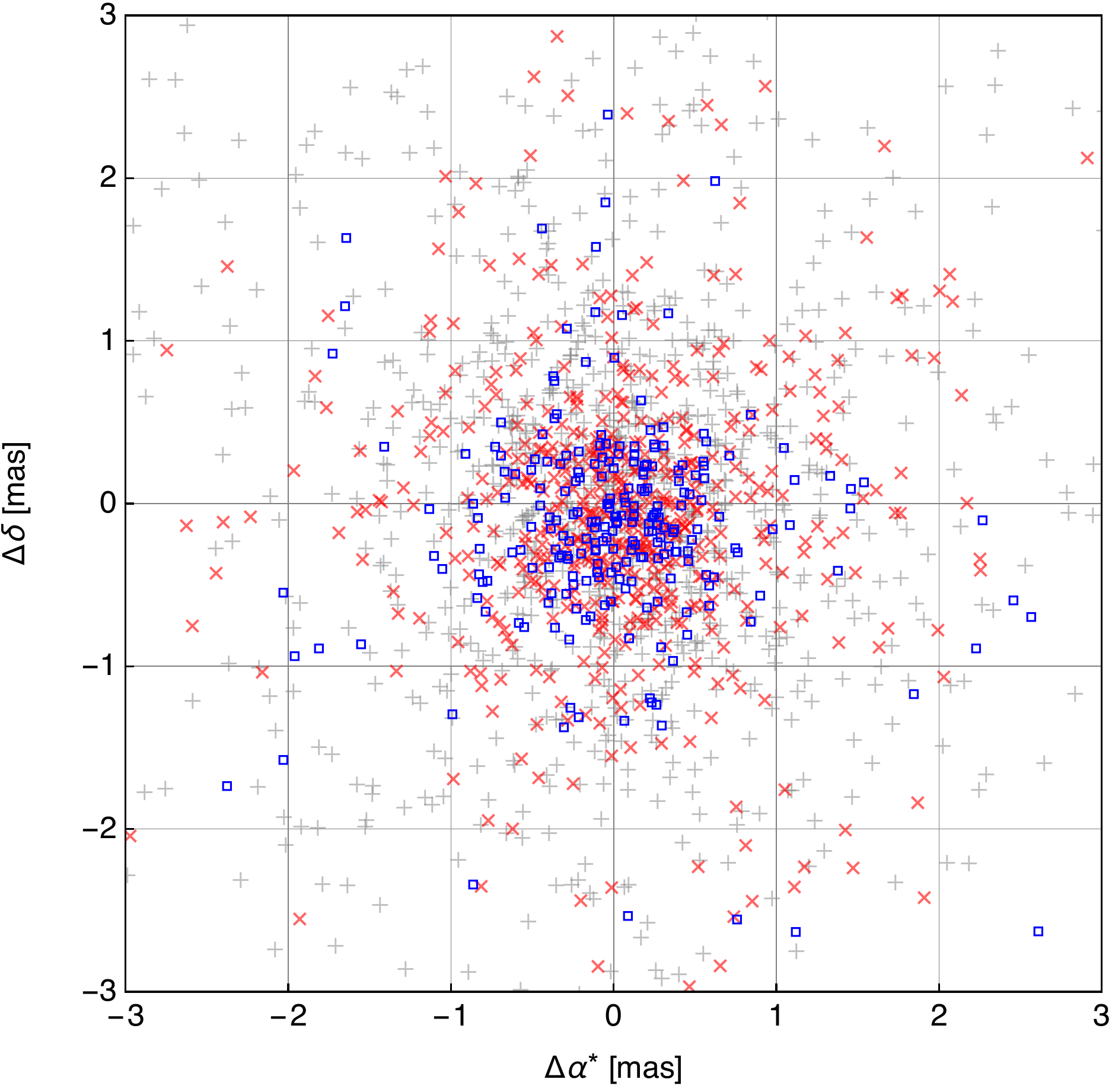}
       \includegraphics[height=0.4\hsize]{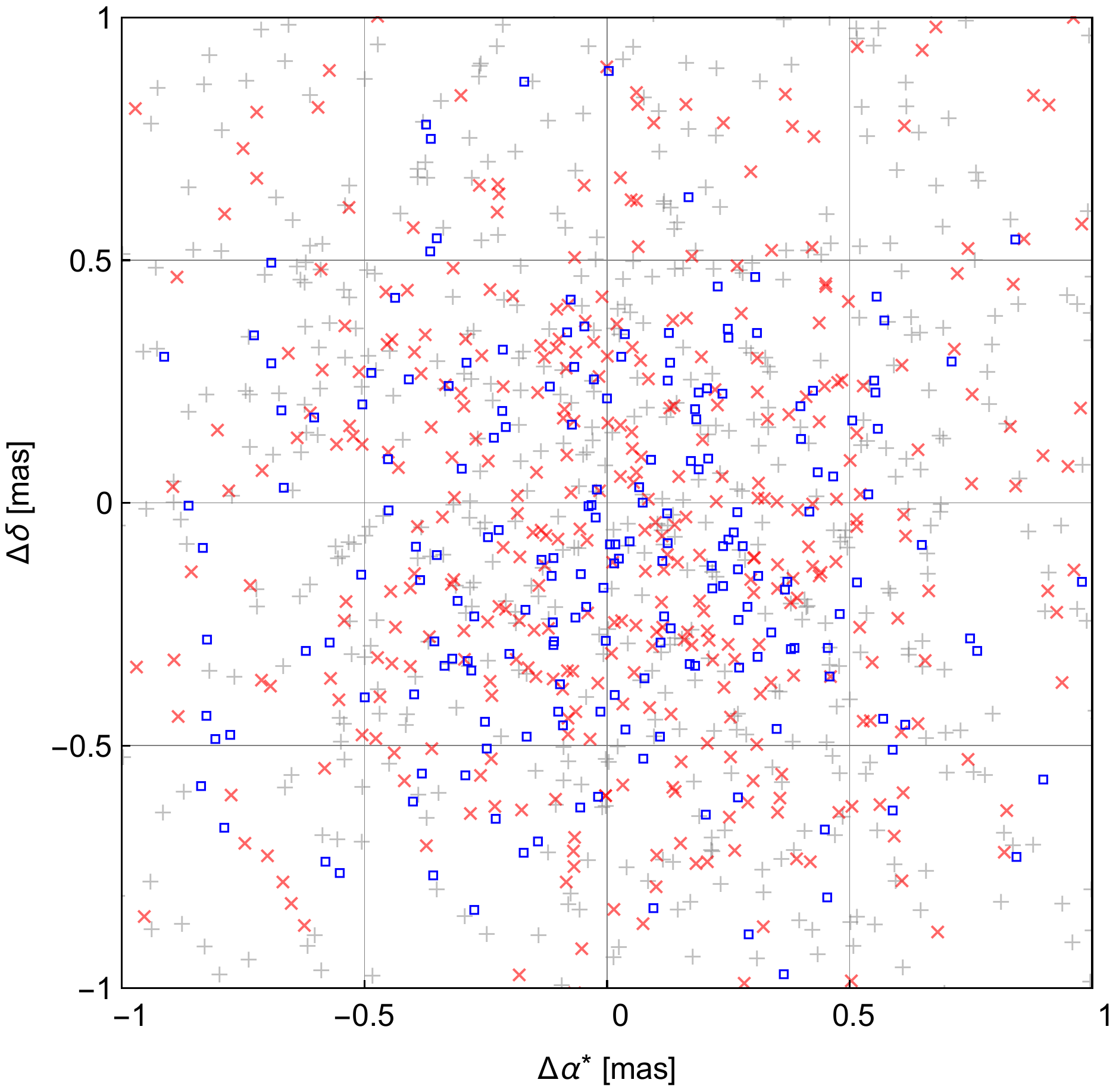}
       }

        \caption{Scatter plot of position differences in right ascension and declination (Gaia minus ICRF2). 
        {\it Top left:} all 2191 sources. {\it Top right:} the 2069 sources within $\pm 10$~mas in both 
        coordinates. {\it Bottom left:} the 1742 sources within $\pm 3$~mas in both coordinates. {\it Bottom right:} the 1069 sources within $\pm 1$~mas in both coordinates. 
        Symbols show defining sources as blue squares, non-VCS sources as red crosses, and
        VCS-only sources as  grey plus signs.\label{fig:da_dd_scatter}}
\end{figure*}

In declination the median $\Delta\delta$ for the whole sample is $-0.069\pm 0.021$~mas,
the RSE 1.85~mas, and the sample standard deviation 7.99~mas. For the defining subset
the median is $-0.136\pm 0.032$~mas, the RSE is 0.70~mas, and the sample standard deviation
3.21~mas. The bias of $\simeq -0.1$~mas in declination is statistically significant and may be 
related to the (ecliptic) north--south asymmetry of systematics noted in Gaia DR1 parallax data
(e.g.\ Appendix~C in \citeads{2016GaiaL}). The reference frame alignment using a solid rotation 
(orientation) cannot remove a possible bias in declination. \footnote{After the submission of the paper the authors have learnt independently from P. Charlot and Ch. Jacobs (private communications) that a similar bias is seen between the ICRF2 and the preliminary solution of ICRF3.}

The plots in Fig.~\ref{fig:da_dd_scatter} show scatter plots of $(\Delta\alpha{*}, \Delta\delta)$, 
colour coded for the different categories of ICRF sources. 
The upper left plot goes up to $\pm 150$~mas in each coordinate and therefore contains all 
2191 sources. It shows primarily the few sources well outside the saturated centre, where 
most of the sources are concentrated with overlapping markers. There are two obvious 
outliers among the defining sources (blue squares); they are discussed in 
Sect.~\ref{sect:poor-sol}. The upper right plot gives the distribution for the 2069 sources
with coordinate differences within $\pm 10$~mas. Here one can clearly see that the defining 
sources, as expected, are more strongly concentrated around the centre than the non-VCS 
sources (red crosses), while the VCS-only sources (grey plus signs) have the largest spread. 
However, there are some 10--15 defining sources outside the central distribution that 
should be examined individually for possible contamination by the host galaxy or other
anomalies. The lower left plot in Fig.~\ref{fig:da_dd_scatter} extending to $\pm 3$~mas shows 
more details near the centre but the distribution remains remarkably symmetric even at the 
highest resolution. The lower right plot in  Fig.~\ref{fig:da_dd_scatter} is a close-up view of the very central region within $\pm 1$~mas in each coordinate  and shows again the predominance of defining sources and the significant fraction of sources in the very inner zone at few 0.1~mas from the centre. The very small bias in declination is now visible with more sources in the negative values of $\Delta\delta$.
The proportion of defining sources (blue squares) increases with resolution, 
but in all three plots the optical and radio positions are statistically in very good agreement
for all three categories of sources despite the larger uncertainties in the radio data for the 
non-defining and especially the VCS-only sources.

\subsubsection{Angular separations}

Table~\ref{table:overall-stat} gives the overall statistics of the angular separation ($\rho$)
between the optical and radio positions, subdivided according to the type of ICRF source. 
It is seen that 94\% of the optical positions are within 10~mas of the radio position of the 
associated ICRF2 source. This is quite remarkable given the limited time coverage and the 
numerous limitations of the astrometric solutions for Gaia DR1 as explained in \citet{2016GaiaL}. 
For the defining sources the percentage is even higher (98\%). There is in fact a very consistent
pattern in that the fraction of separations within a given limit ($\rho_\text{max}$) is always
higher for the defining sources than for the non-VCS sources, which in turn have a higher
fraction than the VCS-only sources. Since the distinction between the three categories is
irrelevant for Gaia (cf.\ Fig.~\ref{spos_gmag_nobs}), this confirms the difference in the quality 
of the radio positions expressed by the quoted uncertainties in ICRF2.

\begin{table}
        \caption{Number of ICRF sources with angular separations $\rho<\rho_\text{max}$ between
        the optical and radio positions. In brackets are the percentages of sources in the group
        below the given limits.\label{table:overall-stat}}
        \small
        \begin{tabular}{lrrrrrr}
	\hline\hline
        \noalign{\smallskip}
ICRF        &  \multicolumn{6}{c}{$\rho_\text{max}$ (mas)} \\ 
group      &       0.5 &         1 &         5  &         10 &        50 &       150  \\
         \hline         
        \noalign{\smallskip}
all            &      451 &     971 &    1884 &     2054 &    2173 &      2191 \\
(\%)          &  (20) & (44) &  (86) &  (94)  &  (99) &  (100) \\[4pt]
defining   &      113 &     186 &      245 &      258  &      261 &        262 \\
(\%)          &  (43) & (71) &  (94) &  (98)  &  (100) &  (100) \\[4pt]
non-VCS  &      167 &     333 &      556 &      599  & 620        & 640      \\
(\%)          &  (26) & (52) &  (87) &   (94)  &  (97) &  (100) \\[4pt]
VCS-only &      171 &     452 &    1083 &     1197  &    1262 &      1289 \\
(\%)          &  (13) & (35) &  (84) &   (93)  &  (98) &  (100) \\
        \hline
        \end{tabular}
\end{table}

\begin{figure*}
       \resizebox{\hsize}{!}{%
       \includegraphics[width=0.4\hsize]{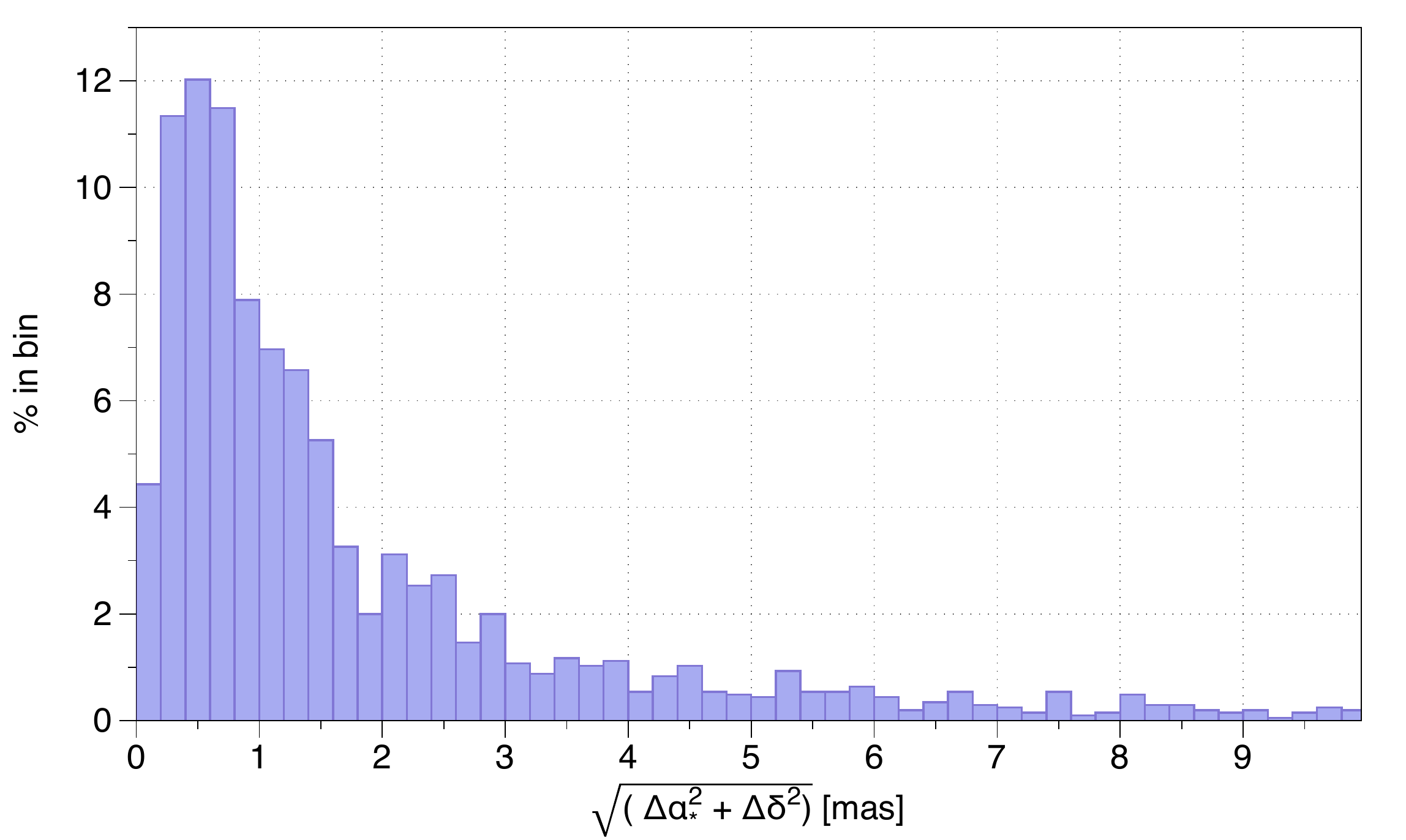}
       \includegraphics[width=0.4\hsize]{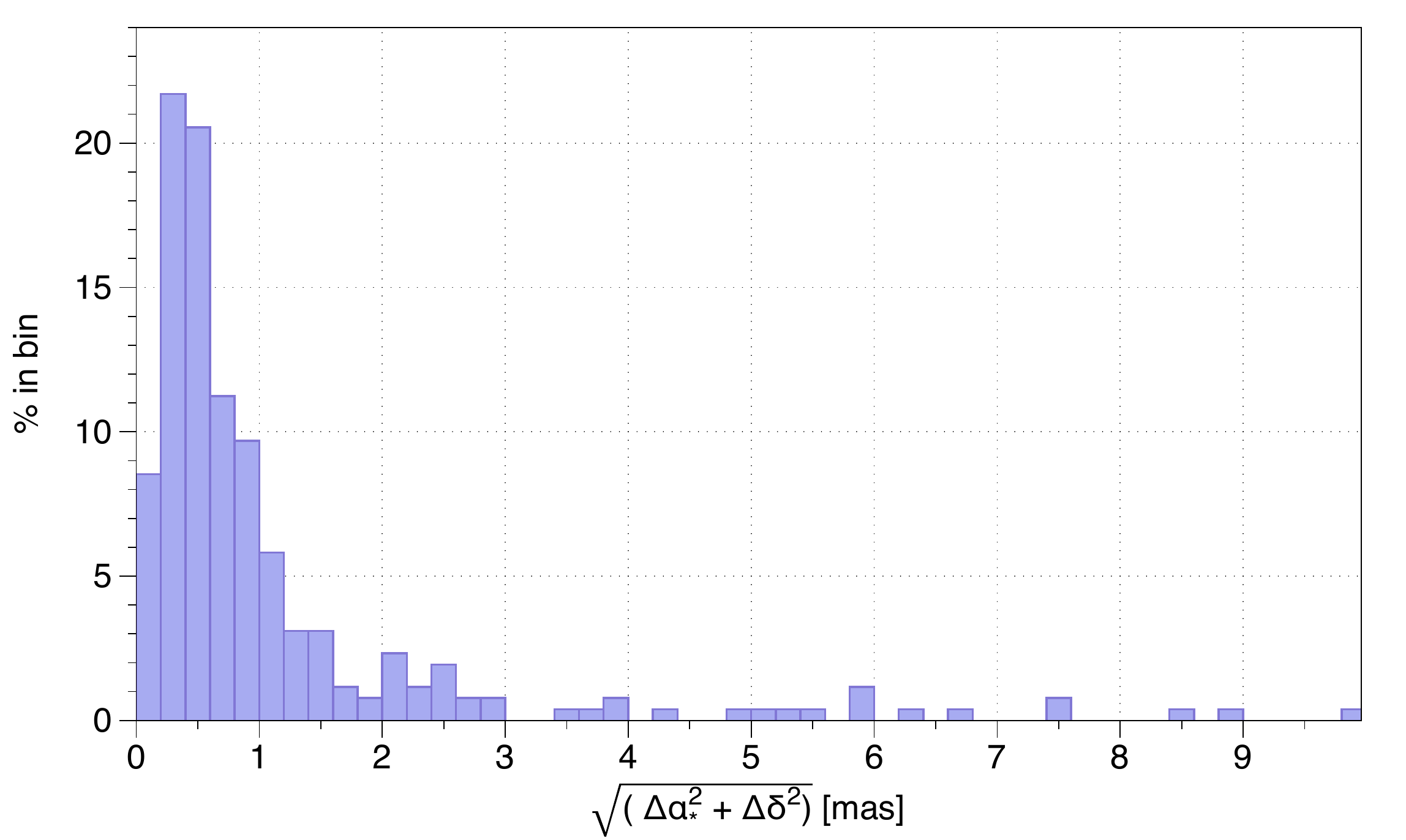}
       }
        \caption{Distribution of the angular distances between the Gaia solution and the ICRF2 positions.   {\it Left:} the 2054 solutions closer than 10 mas to their radio position.  {\it Right:}  same for the  258 defining sources.
        \label{rho_10mas_histo}}
\end{figure*}

Figure~\ref{rho_10mas_histo} shows the distribution of separations in graphical form. 
The difference between the whole set and the set of defining sources is even more conspicuous 
than for the coordinate differences. The median is 1.16~mas in one case and only 0.61~mas in 
the other. As mentioned earlier this is primarily due to the very different quality in the ICRF2 
between the defining sources and the rest of the catalogue and confirms a well known feature of the ICRF2 with fully independent observations.

\subsection{Normalised differences} \label{subsec:norm-compar}

In this section we discuss the positional differences scaled by their standard uncertainties. 
We define the normalised coordinate differences as
\begin{equation}\label{eq:normDiff}
X_\alpha  = \frac{\Delta\alpha{*}}{\sqrt{\sigma^2_{\!\alpha{*},\,\text{Gaia}} 
+\sigma^2_{\!\alpha{*},\,\text{ICRF}}}} \, , \quad
X_\delta  = \frac{\Delta\delta}{\sqrt{\sigma^2_{\!\delta,\,\text{Gaia}} 
+\sigma^2_{\!\delta,\,\text{ICRF}}}} \, ,
\end{equation}
where the ICRF values are taken from the ICRF2 catalogue. The scaled analogue of $\rho$
will be defined later (Sect.~\ref{sect:scaledSep}).

\begin{figure*}[ht]
       \resizebox{\hsize}{!}{%
       \includegraphics{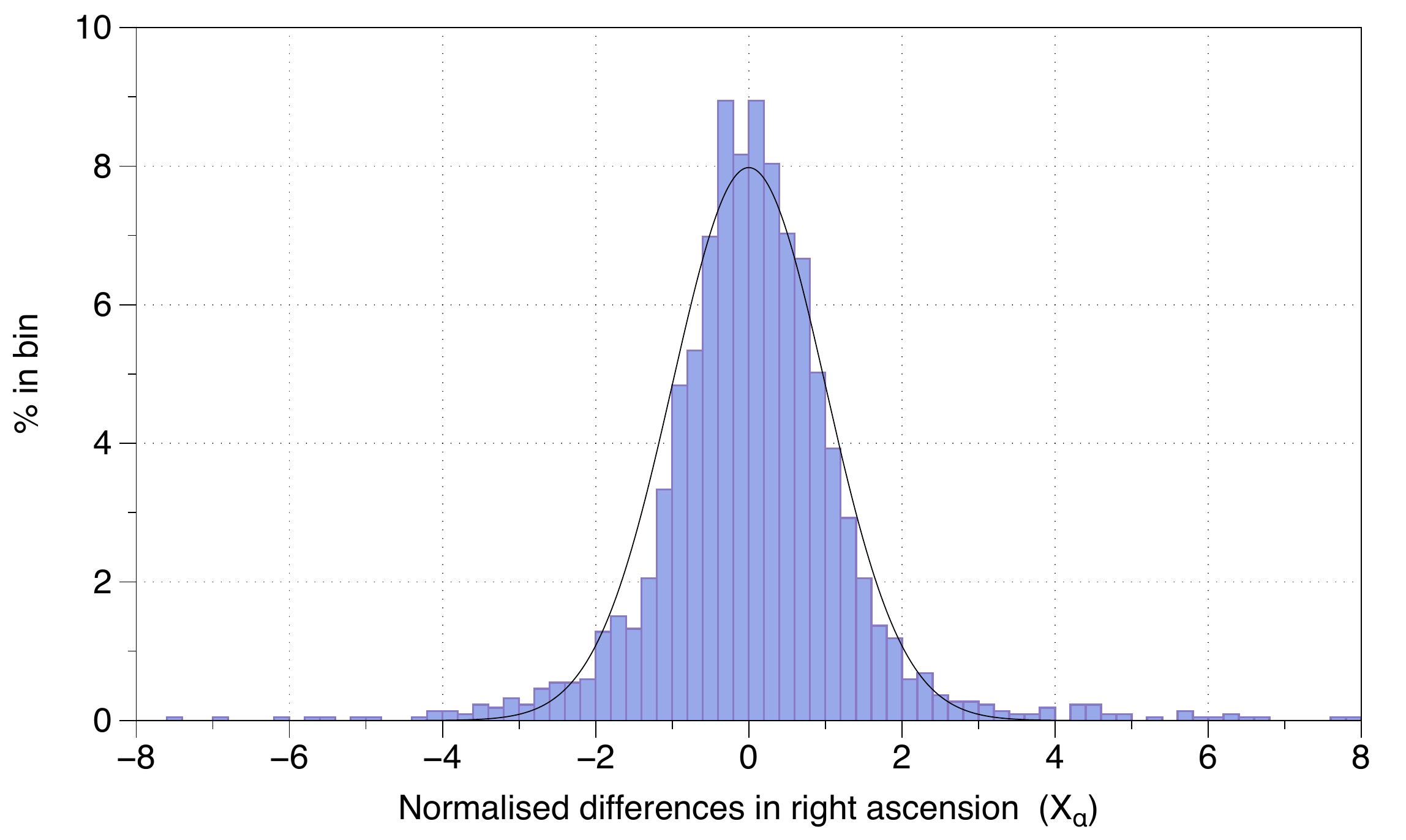}
       \includegraphics{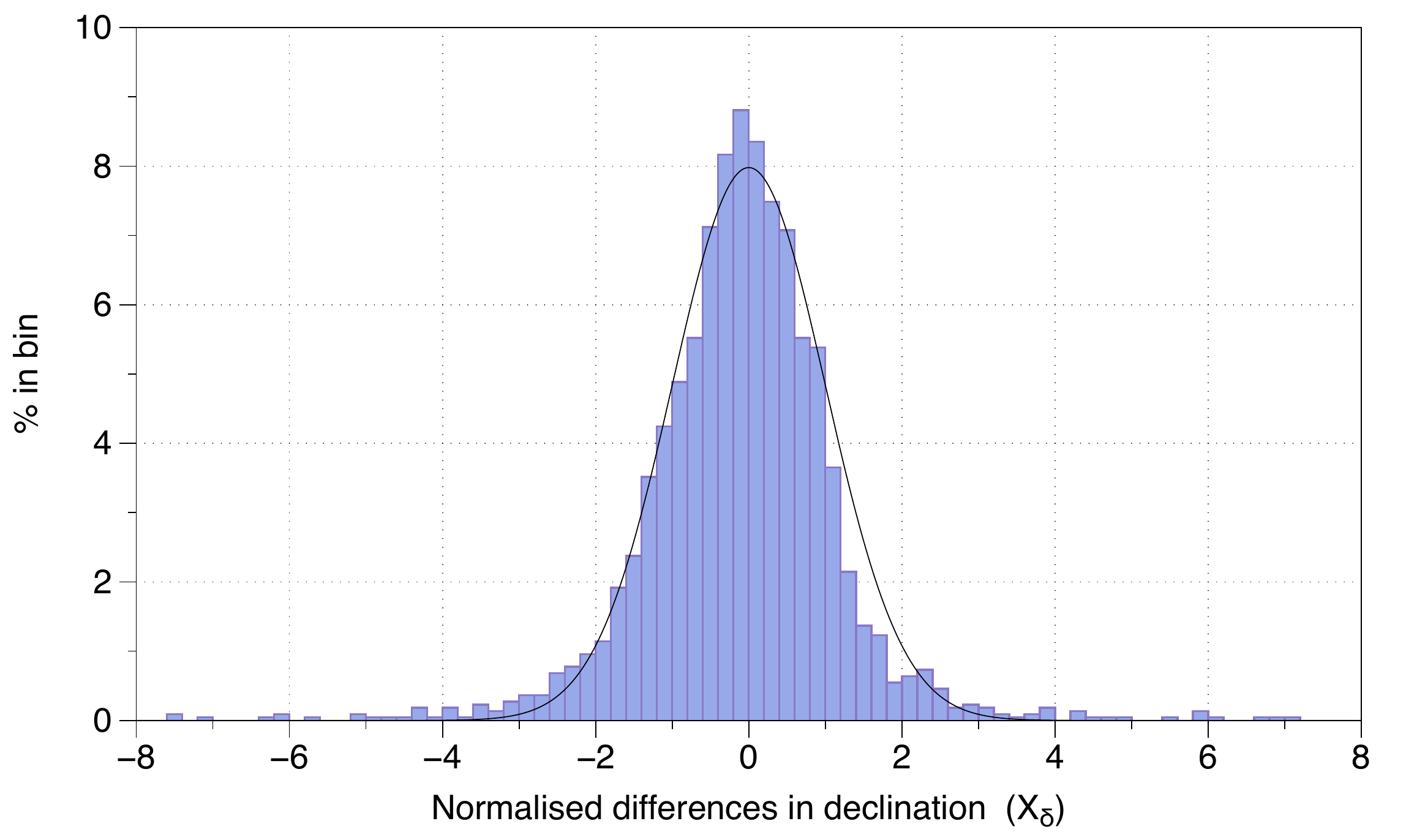}
       }
        \caption{Distribution of the normalised differences in right ascension (left) and declination (right) 
        between the Gaia DR1 positions of 2191 ICRF sources and their radio positions. The histograms are cut 
        at $\pm 8$ in the normalised differences, leaving out 28 sources with $|X_\alpha|>8$ 
        and 28 sources with $|X_\delta|>8$. The black curves are the expected centred normal distribution 
        of unit standard deviation. 
        \label{norm2_da_dd_histo}}
\end{figure*}

\begin{figure*}[ht]
       \resizebox{\hsize}{!}{%
       \includegraphics{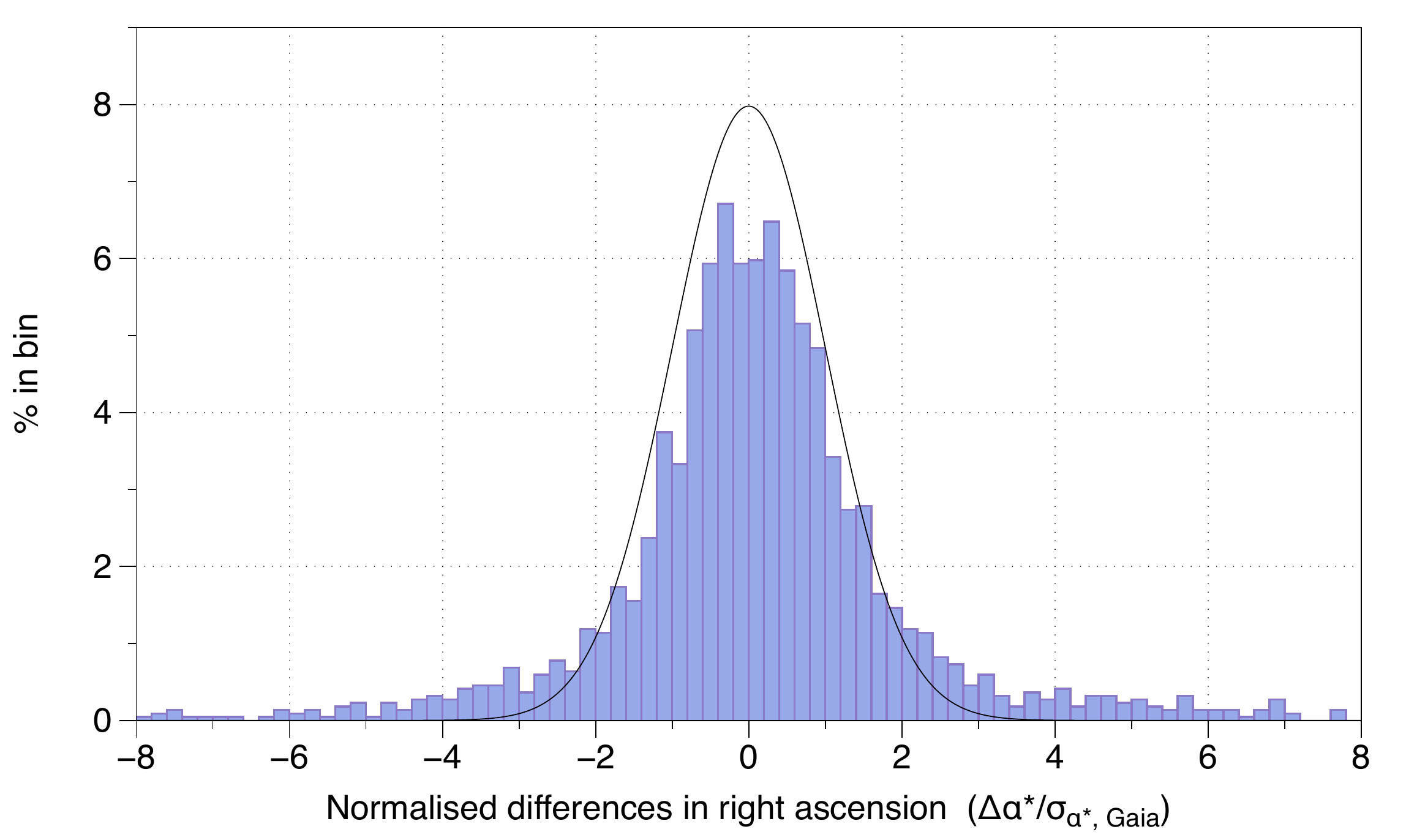}
       \includegraphics{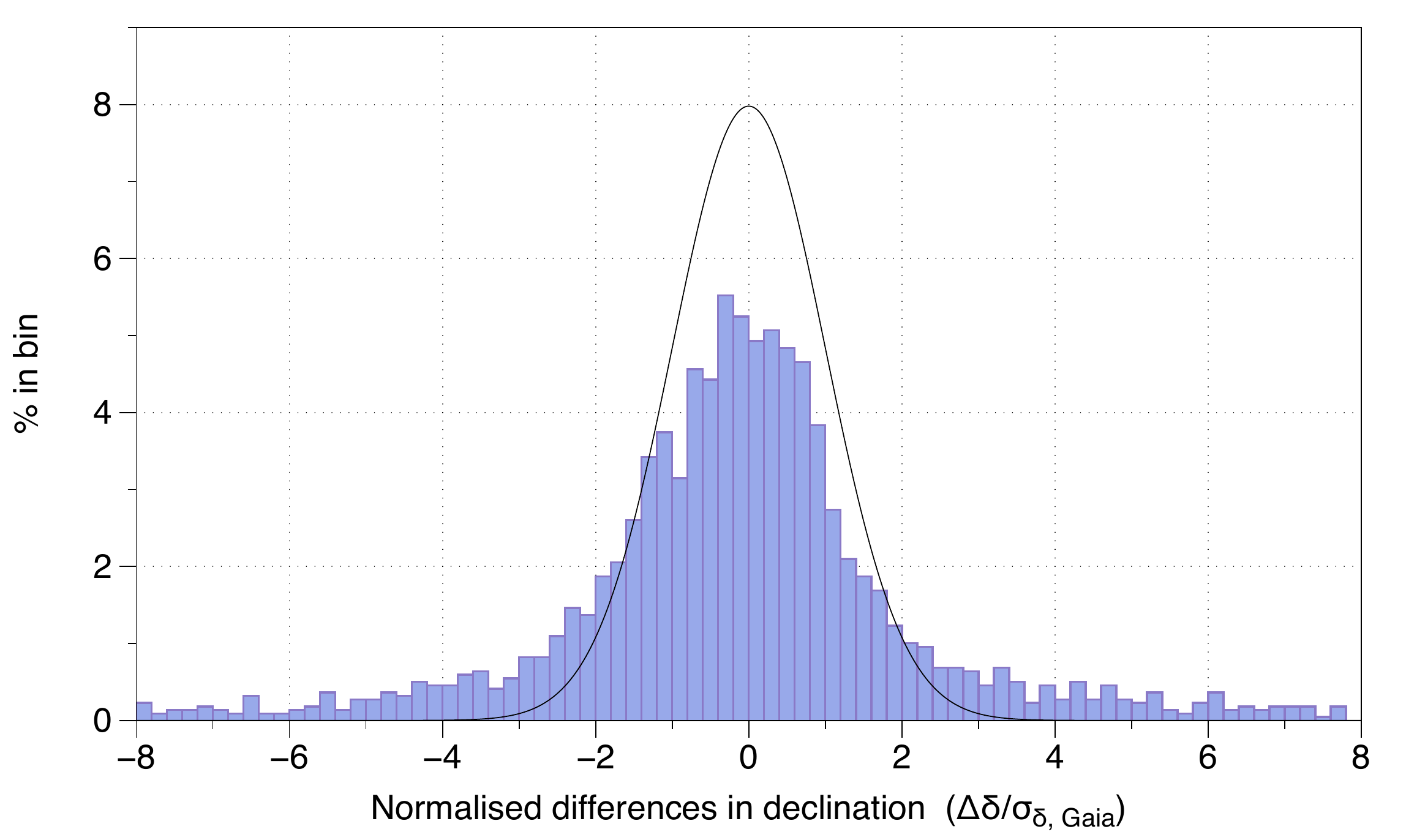}
       }
        \caption{Distribution of the differences in right ascension (left) and declination (right) 
        between the Gaia DR1 positions of 2191 ICRF sources, normalised by the uncertainties of the Gaia data.
        The histograms leave out 97 sources with $|\Delta\alpha{*}/\sigma_{\alpha*}|>8$ 
        and 154 sources with $|\Delta\delta/\sigma_\delta|>8$. The black curves are the expected 
        centred normal distribution of unit standard deviation. 
        \label{norm1_da_dd_histo}}
\end{figure*}

\subsubsection{Normalised coordinate differences}

Figure~\ref{norm2_da_dd_histo} shows the cores of the distributions of $X_\alpha$ and $X_\delta$, 
leaving out only a few dozen points beyond the histogram boundaries. The histograms roughly follow
the expected normal distribution of unit width, although there are clearly several tens of outliers in
both histograms, and the negative bias in declination is clearly visible in the right histogram.
The overall agreement is confirmed by the RSE, which is 1.01 in right ascension and 1.02 in
declination. The outliers increase the sample standard deviations to 2.4 and 2.6, respectively.

The normalisation factors in Eq.~(\ref{eq:normDiff}) include the contributions from the uncertainties
in both data sets (Gaia and ICRF). For comparison we show in Fig.~\ref{norm1_da_dd_histo} the
corresponding plots when only the uncertainties in the Gaia data are included (obtained by setting
$\sigma_{\!\alpha{*},\,\text{ICRF}}=\sigma_{\!\delta,\,\text{ICRF}}=0$ in Eq.~\ref{eq:normDiff}).
As expected, the resulting distributions are significantly wider (the RSE is 1.70 in $\alpha$ and 
2.50 in $\delta$), but the main difference is the much larger number of sources in the 
wings than when the combined uncertainties are used to scale the differences. These are primarily
due to the non-defining sources, for which the radio positions are often much more uncertain than
the optical positions. The good overall agreement when the combined uncertainties are used 
suggests that the uncertainties quoted in ICRF2 for the non-defining sources are fairly reliable.  
Actually the true positional uncertainty of the VCS-only sources is probably difficult to ascertain 
with VLBI observations performed in a single session. The on-going work to prepare ICRF3 with 
repeated observations of VLBA calibrators will be very useful in this respect \citepads{2016AJ....151..154G}.

\begin{figure*}[ht]
       \resizebox{\hsize}{!}{%
       \includegraphics{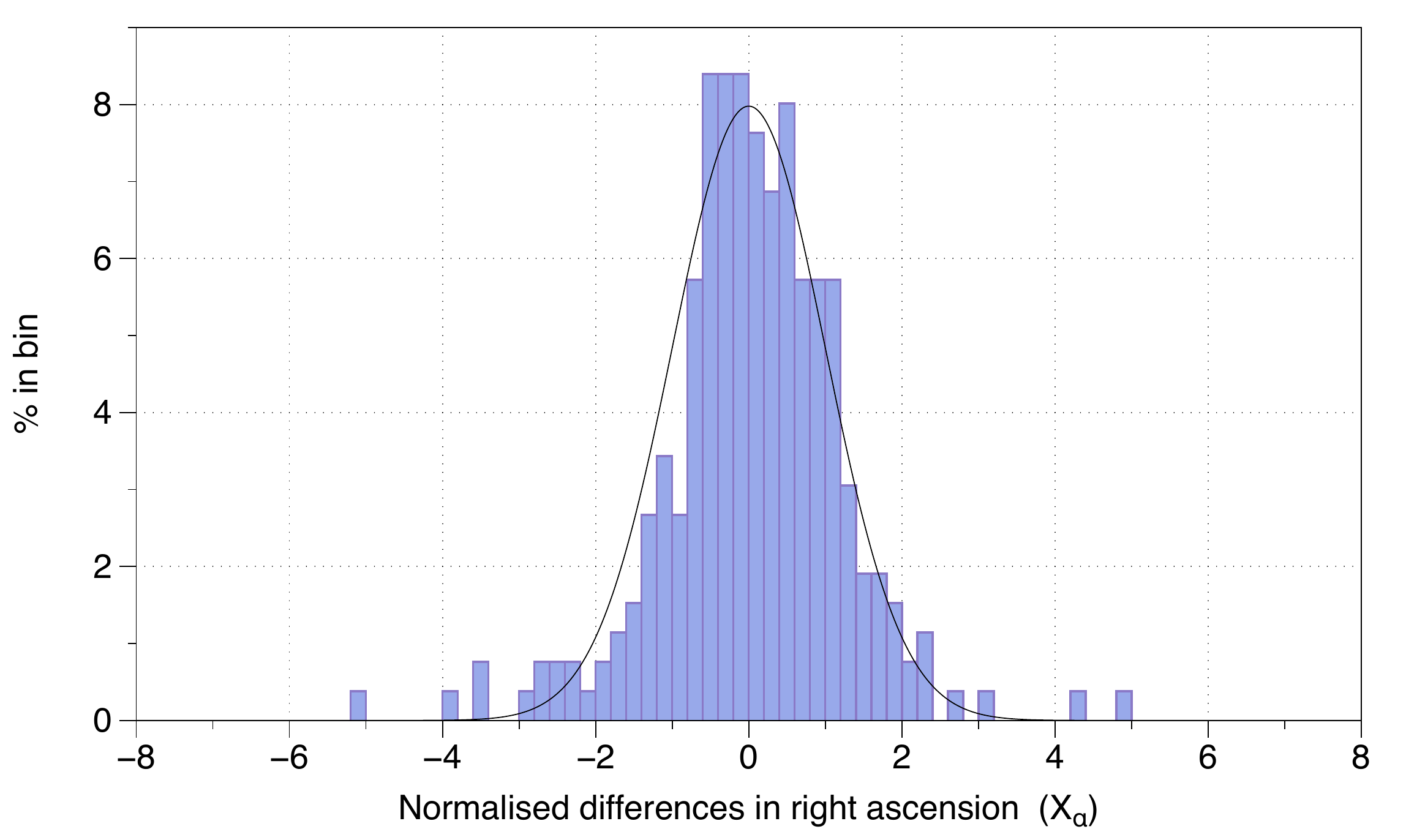}
       \includegraphics{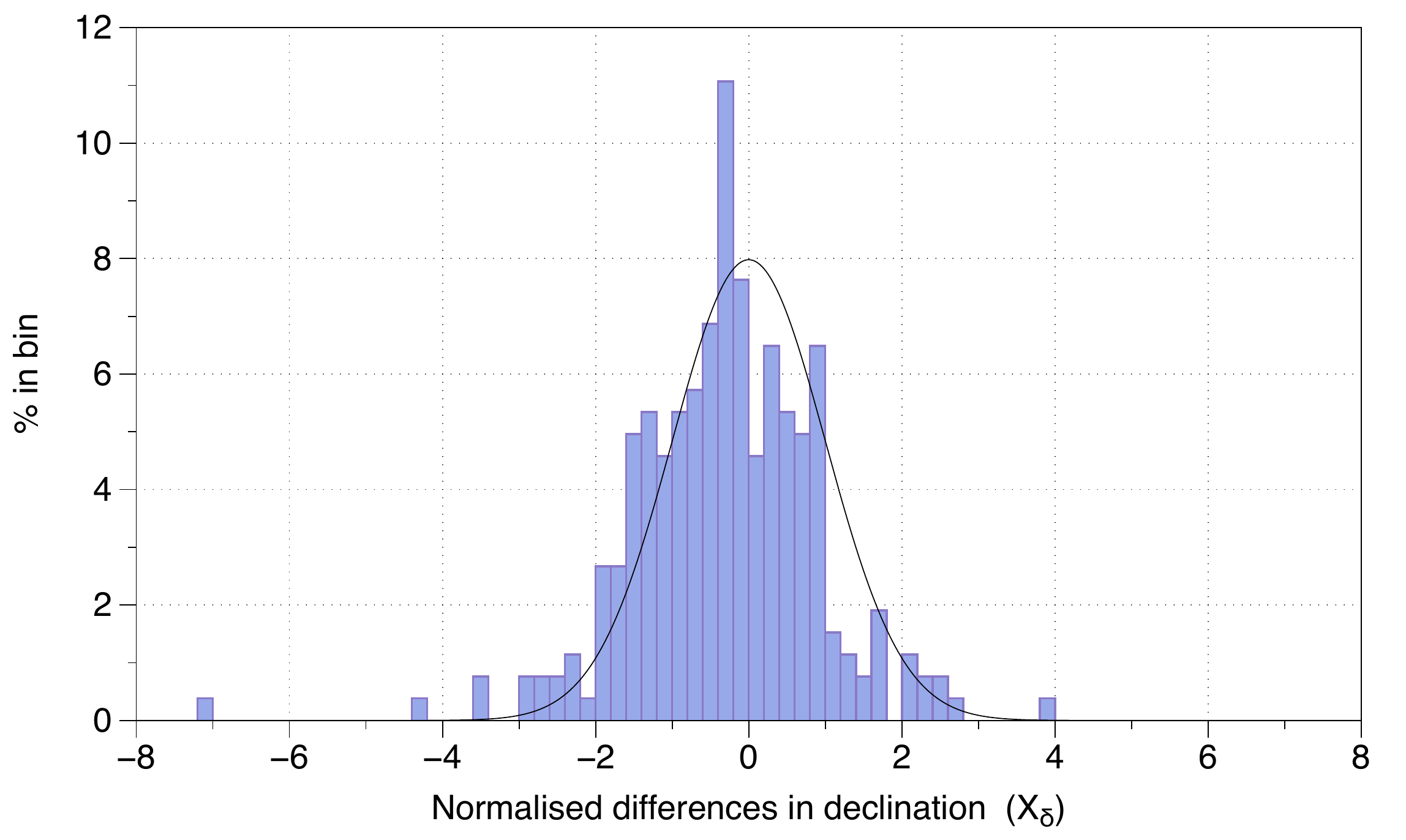}
       }
        \caption{Distribution of the normalised differences in right ascension (left) and declination (right) 
        between the Gaia DR1 positions of 262 defining ICRF sources and their radio positions. 
        The histograms are cut 
        at $\pm 8$ in the normalised differences, leaving out two sources with $|X_\alpha|>8$ 
        and three sources with $|X_\delta|>8$. The black curves are the expected centred normal distribution 
        of unit standard deviation. 
        \label{norm2_da_dd_def_histo}}
\end{figure*}

Focusing now on the defining sources, we show in Fig.~\ref{norm2_da_dd_def_histo} the
distributions $X_\alpha$ and $X_\delta$ for the subset of 262 defining sources in the 
auxiliary quasar solution. For these sources the uncertainty coming from the ICRF2 is negligible, 
with a median around 0.06~mas, in comparison to the Gaia uncertainty, which has a median 
around 0.5~mas. The combined uncertainties in Eq.~(\ref{eq:normDiff}) are therefore dominated 
by the uncertainties from Gaia, and the distribution of the normalised differences is primarily a 
test of the realism of the Gaia uncertainties and of possible physical optical--radio offsets. 
The RSE of the normalised differences is 1.03 in right ascension and 1.05 in declination. The 
deviations of these values from 1.00 are not statistically significant: the uncertainties in RSE
estimated by bootstrap are about 0.08. Considering the much smaller sample, the overall 
agreement with a normal distribution is about as good for the defining sources in 
Fig.~\ref{norm2_da_dd_def_histo} as for the whole sample in Fig.~\ref{norm2_da_dd_histo}. 
The proportion of outliers is also similar, with about 5\% defining sources beyond three standard 
deviations in either coordinate; the corresponding number for the whole sample is about 7\%. 
The bias in declination is more apparent for the defining subset because of the smaller 
combined uncertainties. 
  
If we disregard the bias in declination and a certain percentage of outliers, the distributions 
of the normalised differences for both defining and non-defining sources are in 
reasonable agreement with a normal distribution with zero mean and unit variance.
The normalising factor  takes only into account  the statistical errors and does not include 
the possible additional noise coming from an actual optical--radio offset. If such an offset 
is present in most of the sources, it will be random in direction and will contribute to the 
observed dispersion of coordinate differences, thus increasing the RSE values. This increase
would be more noticeable for the sources with small combined uncertainties. 
From the plots, despite the small number of sources usable for 
this purpose, one can already state that if such offsets exist in the majority of the defining
sources, they must be much less than a mas. The RSE values of 1.03--1.05 for the normalised
differences of the defining sources, although insignificantly larger than one, set an upper limit 
of about 0.4~mas for the optical--radio offsets, since otherwise the RSE values would
be significantly larger than one. This is a statistical result, and does not preclude that the 
offsets are usually small and large only for a small fraction of the sources. By a similar 
reasoning one can conclude that the stated uncertainties of the Gaia positions in general 
cannot be underestimated, in the quadratic sense, by more than about 0.4~mas.

\begin{figure*}[ht]
       \resizebox{\hsize}{!}{%
       \includegraphics{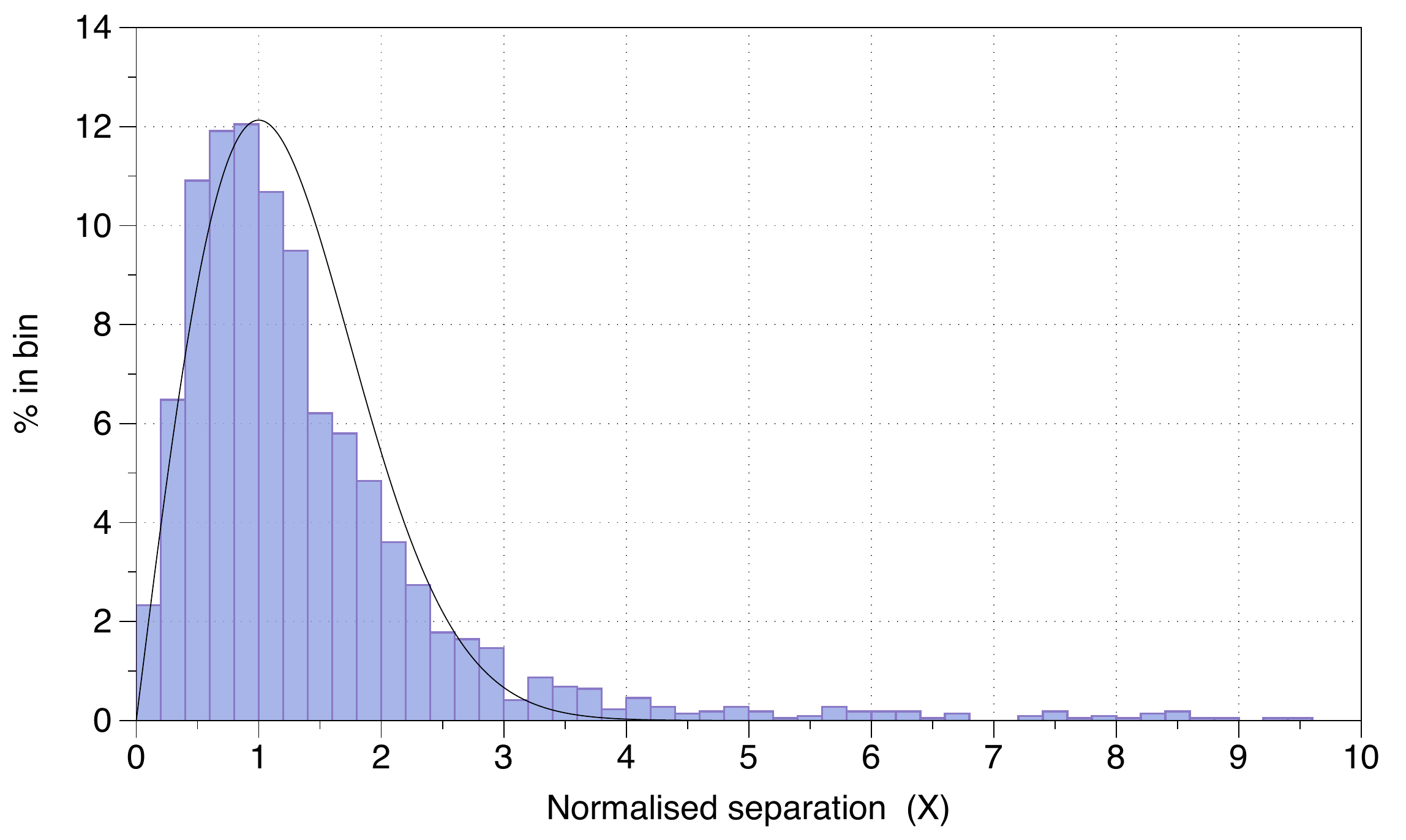}
       \includegraphics{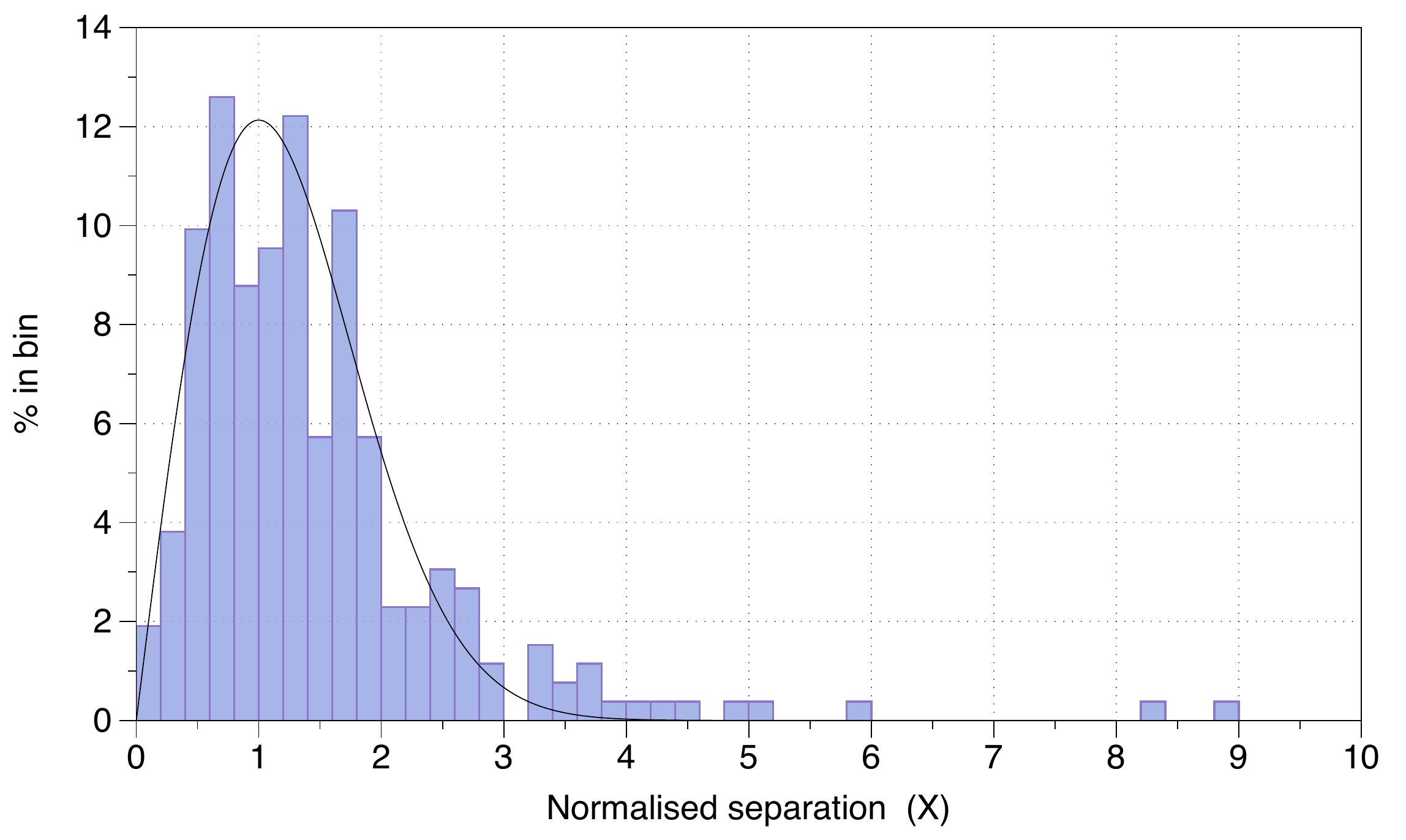}
       }
        \caption{Distribution of the normalised separations $X$ (Eq.~\ref{eq:normSep}) 
        between the Gaia DR1 positions of ICRF sources and their radio positions. 
        {\it Left:} all 2191 sources (36 have $X>10$). {\it Right:} 262 defining sources
        (three have $X>10$). The black curve is the expected Rayleigh distribution. 
        \label{normSep_histo}}
\end{figure*}

\subsubsection{Normalised separations}\label{sect:scaledSep}

When considering the normalised differences in both coordinates jointly, the obvious 
analogue of $\rho$ in Eq.~(\ref{eq:rho}) would be $(X_\alpha^2+X_\delta^2)^{1/2}$.
This quantity can however be misleading when there are strong correlations between
the errors in $\alpha$ and $\delta$, which is often the case especially for the Gaia data.
Indeed, for Gaussian errors the theoretical distribution of this quantity will depend on 
the degree of correlation between the two coordinates. To take into account the correlation 
coefficients $C_\text{Gaia}$ and $C_\text{ICRF}$ in both data sets, we use the statistic
\begin{equation}\label{eq:normSep}
X^2 =  \begin{bmatrix} X_\alpha & X_\delta \end{bmatrix}
\begin{bmatrix} 1 & C \\ C & 1 \end{bmatrix}^{-1}
\begin{bmatrix} X_\alpha \\  X_\delta \end{bmatrix} \, ,
\end{equation}
where
\begin{equation}\label{eq:normC}
C =  \frac{\sigma_{\alpha{*},\,\text{Gaia}}\sigma_{\delta,\,\text{Gaia}}C_\text{Gaia}
+\sigma_{\alpha{*},\,\text{ICRF}}\sigma_{\delta,\,\text{ICRF}}C_\text{ICRF}}%
{\sqrt{\left(\sigma^2_{\!\alpha{*},\,\text{Gaia}}+\sigma^2_{\!\alpha{*},\,\text{ICRF}}\right)
\left(\sigma^2_{\!\delta,\,\text{Gaia}}+\sigma^2_{\!\delta,\,\text{ICRF}}\right)}}
\end{equation}
is the correlation coefficient of the combined errors. For Gaussian errors we expect 
$X^2$ to follow the chi-squared distribution with two degrees of freedom (see Appendix \ref{sect:deriveqRayleigh} for the mathematical details). Equivalently, 
the normalised separation $X$ should have the Rayleigh distribution, that is
$\text{Pr}(X > x)=\exp(-x^2/2)$. Figure~\ref{normSep_histo} shows the distributions
of $X$ for all the ICRF sources and for the defining sources. The median value of $X$
is $1.11\pm 0.02$ for the whole sample of 2191 sources, and $1.28\pm 0.05$ for the 
defining subset. Compared with the theoretical value $(\ln 4)^{1/2}\simeq 1.18$, the
dispersion is slightly too large for the defining sources and too small for the whole
sample. The decreasing trend is consistent when going from defining to non-VCS and 
VCS-only sources, with median values 1.28, 1.26, and 1.03, respectively. Since this trend
is opposite to the variation of the formal uncertainties in ICRF2, it could be explained by 
a constant contribution from actual optical--radio offsets, or by the formal uncertainties 
in ICRF2 being overestimated for the VCS-only sources and possibly underestimated
for the other sources. That the root cause of this trend is in the Gaia data seems less
likely, as the three categories of ICRF sources have similar properties in the Gaia observations.
The larger-than-expected median values for the defining and non-VCS sources could still
be explained by an underestimation of the Gaia uncertainties, in which case the ICRF2
uncertainties of the VCS-only sources would have to be even more overestimated. 

Concerning the tail of large normalised separations we note that, 
for the theoretical distribution, the expected number of points with $X>4.1$ 
in a sample of size 2191 is less than 0.5. In reality we find 107 sources 
above this limit, including 11 defining sources. These can reasonably be considered as 
outliers requiring a more detailed analysis.

\begin{figure}[h]
       \resizebox{\hsize}{!}{%
       \includegraphics{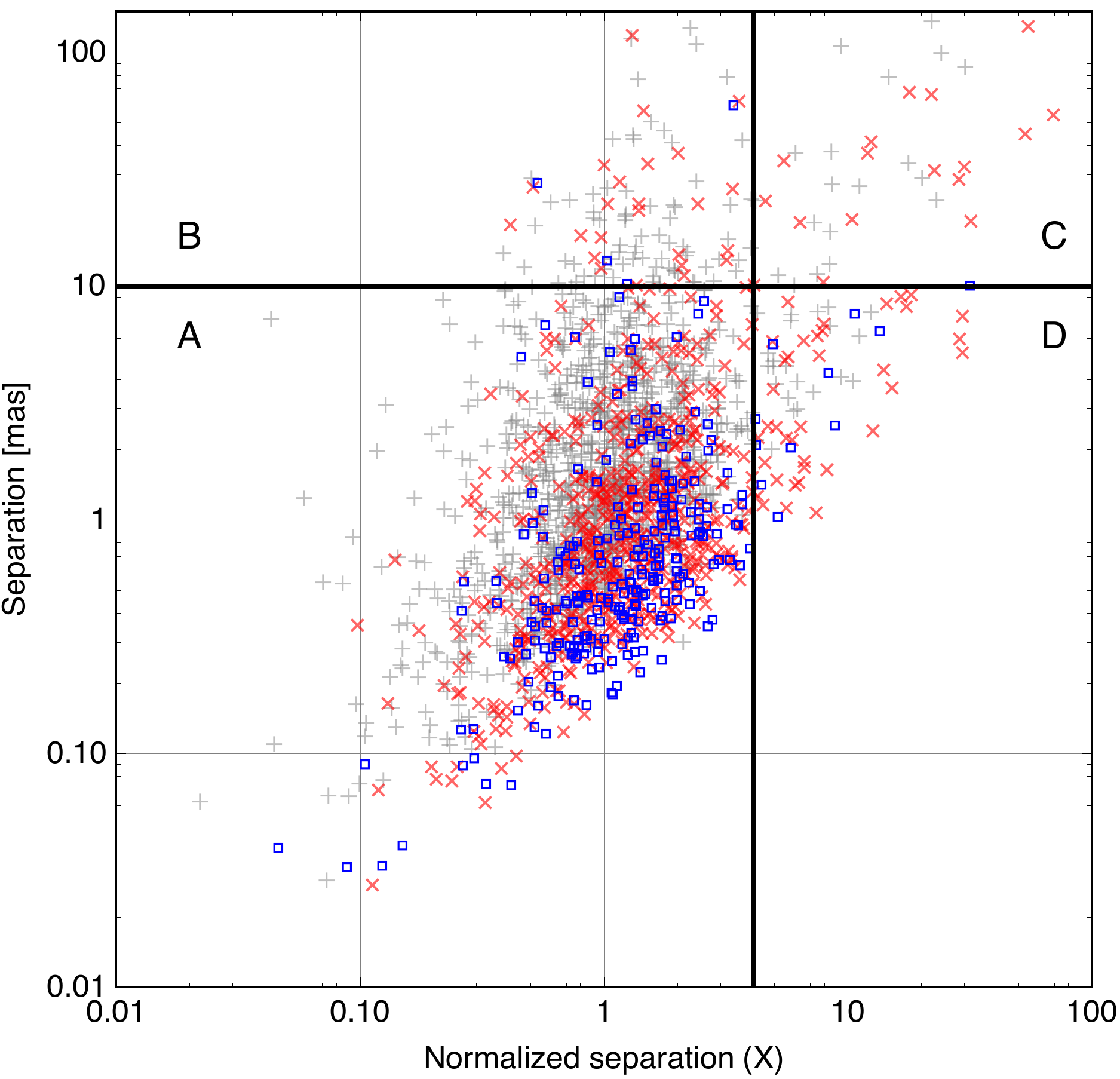}
       }
        \caption{Angular separation $\rho$ versus normalised separation $X$ for 2191 ICRF sources.
         Defining sources as shown as blue squares, non-VCS sources as red crosses, and
        VCS-only sources as pale grey plus signs.
        The four regions labelled A, B, C, D are separated by $\rho=10$~mas and $X=4.1$. 
        \label{poor_sol_scat}}
\end{figure}

\subsection{Deviating cases}\label{sect:poor-sol}

We now consider the sources whose optical positions in the Gaia data deviate 
markedly from the radio positions of the matched ICRF sources, either in angular 
separation ($\rho$) or in normalised separation ($X$). Figure~\ref{poor_sol_scat} 
is a plot of $\rho$ versus $X$ for all 2191 sources. In Sect.~\ref{sect:scaledSep} 
we concluded that sources with $X>4.1$ have a larger separation than expected 
from the combined formal uncertainties, and can
therefore be considered as deviating cases on purely statistical grounds. This 
subset contains 107 sources, or 5\% of the whole sample.

The sources with atypically large angular separations are however also interesting,
for reasons more related to the physical properties of the sources, their environment 
on the sky, and the resolution of the Gaia instrument. Here it is much harder to 
motivate a~priori any specific limit in angular separation beyond which a source merits 
special attention. Somewhat arbitrarily, we choose to look at the sources with 
$\rho>10$~mas, a value significantly larger than the typical combined uncertainty. 
There are 137 sources that are singled out by this criterion, a number small enough 
to think of a case-by-case investigation.

The limits in $X$ and $\rho$ divide the 2191 ICRF sources into four categories,
corresponding to the regions labelled A, B, C, and D in Fig.~\ref{poor_sol_scat},
and which are treated separately below. It has not been possible to investigate
every deviating source; only a few representative or extreme cases are discussed
in some detail. Individual cases have been examined using on-line resources
available through the Strasbourg astronomical Data Center (CDS), including 
the Sloan Digital Sky Survey 
(SDSS; \citeads{2000AJ....120.1579Y}, \citeads{2011AJ....142...72E})
and the the Two Micron All Sky Survey (2MASS; \citeads{2006AJ....131.1163S}).
\begin{enumerate}[A.]
\item  % region "A"
This is the unproblematic category where both $\rho$ and $X$ are unremarkable.
It contains 1982 sources (247 defining, 562 non-VCS, and 1173 VCS-only sources),
which will not be further discussed.
\item  % region "B"
In this category the angular separation exceeds 10~mas, but $X$ is still below
the limit which means that the large separation may not be statistically significant.
This category contains 102 sources (4 defining, 24 non-VCS, and 74 VCS-only sources).
The four defining sources have been examined individually. In order of decreasing
angular separation they are:
\begin{itemize}
\item \object{ICRF J121546.7-173145} ($\rho=58.9$~mas) is within 40\arcsec\ of the 
2nd magnitude star $\gamma$~Corvi. This produces a large glare around the image, 
increasing the background noise and generating many false detections. It is possible
that Gaia will never deliver a good solution for this source.
\item \object{ICRF J173302.7-130449} ($\rho=27.3$~mas) is at 30\arcsec\ from a star of
magnitude 7.5. It is not obvious how much trouble this produced in the Gaia detection, 
but the observation record shows that the detections were not always matched to 
the same Gaia source, which supports the conjecture that false detections have 
contaminated the solution.
\item \object{ICRF J144553.3-162901} ($\rho=12.8$~mas) is a rather poorly observed source with 
only eight matched detections and a positional uncertainty around 15~mas in the Gaia data. 
The optical source is isolated in the SDSS and 2MASS surveys.
\item
\object{ICRF J023945.4-023440} ($\rho=10.1$~mas) is also a poorly observed source with  
seven matched detections and Gaia uncertainties around 8~mas. Again, the optical source 
is isolated in the SDSS and 2MASS surveys.
\end{itemize}
The most deviating non-VCS source in this category is ICRF J054138.0$-$054149
($\rho=117.9$~mas, $X=1.3$). The auxiliary quasar solution has large uncertainties
for this source, 
although there is a good number of matched detections. No nearby bright star or host galaxy 
is visible in the 2MASS or SDSS survey, but the source is very faint for Gaia at $G= 20.7$.

\smallskip
The Gaia uncertainties for these five sources are characterised by very elongated error
ellipses (axis ratios in the range 9 to 185) caused by a poor coverage in terms of the scanning
directions. This will definitely improve when more observations are used in the solution
and at least the last two defining sources should then get good positions from Gaia. 
\item  % region "C"
This category contains sources with angular separations exceeding 10~mas 
that are statistically significant in relation to the formal uncertainties. This is where we
are most likely to find a clear reason for the offset of the optical centre from the radio
centre, caused for example by the  host galaxy or a nearby faint star. The category 
includes 35 sources (17 non-VCS and 18 VCS-only sources), with one defining source 
just below the separation limit (see under D below). In order of decreasing $X$ the 
four non-VCS sources with the highest significance are:
\begin{itemize}
\item
\object{ICRF J013741.2+330935} ($X=70$, $\rho=54$~mas) is 3C~48 embedded in 
the extended galaxy \object{SDSS9 J013741.30+330935.0}.
\item
\object{ICRF J133037.6+250910} ($X=55$, $\rho=128$~mas) is 3C~287. 
On SDSS there are two optical objects of 
comparable magnitude (around 18) at 3\arcsec\ from each other, with the ICRF position 
roughly coinciding with the brighter object. From the SDSS images it is difficult to
determine if this object is extended. The optical position from Gaia is offset from
the radio position by 128~mas towards north-east, away from the other object, but
it is difficult to see how such a faint object at 3\arcsec\ could perturb the Gaia measurements.
\item
\object{ICRF J120321.9+041419} ($X=53$, $\rho=45$~mas) is at the centre of a weak
extended galaxy image, \object{SDSS9 J120321.93+041419.0}, larger than 1\arcsec\ on 
SDSS images.
\item
\object{ICRF J114722.1+350107} ($X=32$, $\rho=19$~mas) is at the centre of an 
extended galaxy image, 10\arcsec\ in diameter but with a luminous core. The image 
has a high background for Gaia but with a bright point-like core, so Gaia performs 
nominally well, but the optical centre could be displaced by the surrounding bright 
distribution of light.
\end{itemize}
The Gaia solutions for these sources are formally good, with uncertainties below 1~mas 
and moderate axis ratios (in the range 1.4 to 4). 
\item  % region "D"
This category contains sources with moderate offsets (1--10~mas) that are nevertheless
significant thanks to the small formal uncertainties in both the optical and radio data.
It contains 72 sources (11 defining, 37 non-VCS, and 24 VCS-only sources). The most
discrepant defining source is \object{ICRF J155850.2-643229} ($X=32$, $\rho=9.97$~mas).
The relatively bright source seen by Gaia ($G=16.3$) could be the central part of the
galaxy \object{2MASX J15585027-6432298} with a total magnitude around 14 in the 
Gaia band. The 10~mas offset corresponds to a shift of the optical centre by only 15~pc 
at the distance of the galaxy, which is not unreasonable if the optical emission 
associated with the radio source is not strong.
\end{enumerate} 
Several of the sources in categories~C and D clearly merit further investigation using
high-resolution imaging.

\begin{figure}[h]
       \resizebox{\hsize}{!}{%
       \includegraphics{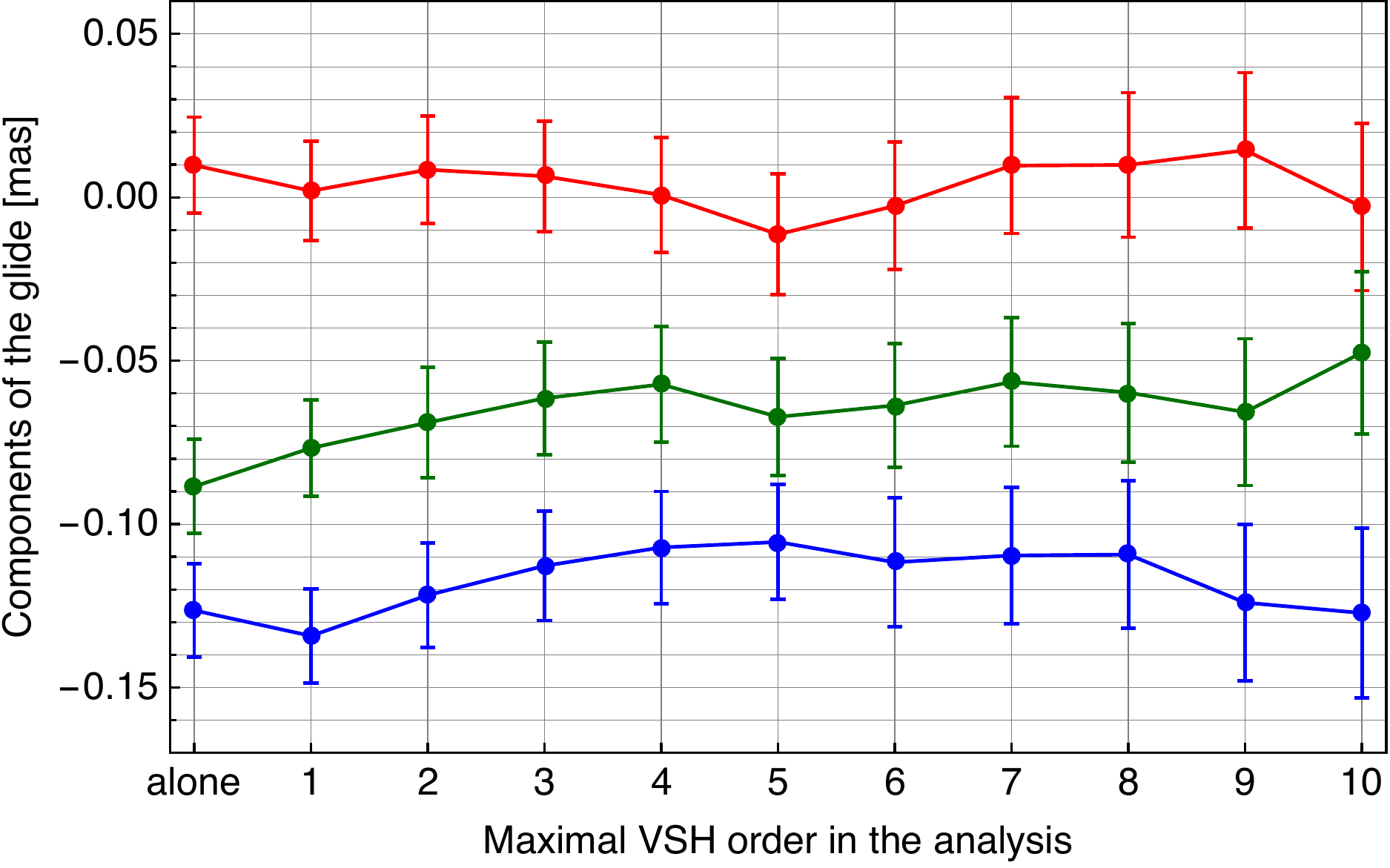}
       }
        \caption{Components of the glide and their errors ($x$, $y$
          and $z$ components are shown in red, green, and blue,
          respectively) for different maximal VSH orders used in
          the fit (``alone'' means that the three glide parameters are
          fitted without the three rotation parameters). 
          The  values correspond to the weighted solutions with all 
          sources in region~A as in Table~\ref{table:alignment}. 
        \label{fig-glide-stability}}
\end{figure}

\begin{table*}[ht]
        \caption{Global differences between the Gaia DR1 positions of ICRF sources and their 
        positions in ICRF2, expressed by the orientation and glide parameters. All numbers are 
        in mas and the VSH analysis was done with $l_{\rm max}=5$.\label{table:alignment}}
        \small
        \begin{tabular}{lcccccccccc}
	\hline\hline
        \noalign{\smallskip}
        & & && \multicolumn{3}{c}{Rotation} &&  \multicolumn{3}{c}{Glide}  \\
        Source selection & Weighting  & $N$ && $x$ & $y$ & $z$ &&  $x$  &  $y$ & $z$         \\
         \hline         
        \noalign{\smallskip} 
all with $\rho<5$~mas  & no &1884  && $\phantom{-}0.00  \pm 0.04$ & $0.03 \pm 0.04$ &$0.01 \pm 0.04$ &&$\phantom{-}0.00 \pm 0.04$ &$ -0.11 \pm 0.04$ &$-0.05 \pm 0.04$ \\
all with $\rho<10$~mas & no &2054  &&            $-0.03 \pm 0.06$ & $0.02 \pm 0.06$ &$0.00 \pm 0.05$ &&$          -0.02 \pm 0.06$ &$ -0.08 \pm 0.06$ &$-0.10 \pm 0.05$ \\
all & yes      &2191  && $\phantom{-}0.02  \pm 0.05$ & $0.06 \pm 0.04$ &$0.04 \pm 0.05$ &&$\phantom{-}0.01 \pm 0.05$ &$ -0.08 \pm 0.05$ &$-0.13 \pm 0.05$ \\
all in region A & yes &1982  && $\phantom{-}0.03  \pm 0.02$ & $0.03 \pm 0.02$ &$0.01 \pm 0.02$ &&$\phantom{-}0.01 \pm 0.02$ &$ -0.07 \pm 0.02$ &$-0.11 \pm 0.02$ \\
defining & yes &\phantom{0}262  && $\phantom{-}0.00  \pm 0.08$ & $0.04 \pm 0.07$ &$0.01 \pm 0.08$ &&$-0.04 \pm 0.07$ &$ -0.14 \pm 0.07$ &$-0.07 \pm 0.07$ \\
defining in region A & yes &\phantom{0}247  && $\phantom{-}0.01  \pm 0.04$ & $0.01 \pm 0.03$ &$0.03 \pm 0.04$ &&$-0.01 \pm 0.04$ &$ -0.12 \pm 0.04$ &$-0.12 \pm 0.04$ \\
        \hline
        \end{tabular}
        \tablefoot{$\rho$ is the angular separation between the optical and radio positions. 
Region A is defined in Sect. \ref{sect:poor-sol} (see Fig. \ref{poor_sol_scat}).
        The weighted solutions use a non-diagonal weight matrix resulting from the combination of Gaia and ICRF2 covariances. In this way not only 
a quadratic combination of Gaia and ICRF2
uncertainties is considered, but also the effective correlation given by Eq. (\ref{eq:normC}). If the correlation is neglected the estimates are typically shifted by a half of the corresponding uncertainty.
$N$ is the number of sources
        used in the solution. The columns headed $x$, $y$, $z$ give the components of the 
        rotation and glide along the principal axes of the ICRS.
        }
\end{table*}

\subsection{Large-scale systematic differences} \label{subsec:alignment}

In this section we compare the two sets of positions in search of general large-scale patterns, 
like a global rotation or a glide, that could explain a fraction of the position differences. 
The positional reference frame of Gaia DR1, based on the auxiliary quasar solution, has been
aligned with ICRF2 as explained in the Gaia DR1 astrometry paper \citep{2016GaiaL}. We therefore
do not expect to see any systematic difference that could be represented by a solid rotation.
The presence of a small difference in declination, noted in Sect.~\ref{sect:coordDiffUnscaled}, does
however suggest other kinds of systematics that should be examined.

Our analysis is based on the expansion of the vector field of the differences (two components 
in the tangent plane of the celestial sphere at the location of each source) on a set of 
vector spherical harmonics (VSH) as explained in \citetads{2012A&A...547A..59M} or 
\citetads{2014MNRAS.442.1249V}. Instead of fitting the differences on a model including the 
rotation, the positional differences are expanded on a set of orthogonal functions up to a 
certain degree $l_\text{max}$. The global rotation and the glide (the vector field locally 
perpendicular to the rotation field; see Sects.~4.2--4.3 of \citeads{2012A&A...547A..59M}) 
are extracted from the harmonics of degree $l=1$. 
This procedure is more general than the alignment since it includes higher-order terms 
that could otherwise project on the rotation when this simplified model is used in isolation. 
This test is therefore an independent check of the initial alignment as well as a search for
potential large-scale systematics.  

The results shown in Table~\ref{table:alignment} are very satisfactory
for the present state of the Gaia data. In particular they confirm the
absence of global rotation at least on the level of 0.05~mas. They do
however indicate a glide of about 0.13~mas, consistent with the
declination difference mentioned above. The values of the glide
parameters are rather stable with respect to the parameters of the fit
as illustrated by Fig.~\ref{fig-glide-stability}. While it is not
possible to tell, from this comparison alone, if the glide results
from a distortion of the Gaia DR1 reference frame or of the ICRF, we
note that the validation of the Gaia DR1 astrometry has revealed many
sorts of systematics at the level of 0.1--0.3~mas
(\citealt{2016GaiaA}, \citealt{2016GaiaL}). It is therefore quite
%likely that this component of the differences originates in the Gaia data. 
possible that the reference frame of Gaia DR1 is distorted by a similar amount.
Apart from the glide, the VSH expansion shows few terms of
higher order (e.g.\ $l = 3$) with significant amplitudes, albeit very
small.

Part of the observed glide component may be caused by the apparent
proper motion of quasars induced by the galactocentric acceleration
of the solar system barycentre (\citeads{1983jpl..rept.8339F},  % Fanselow
\citeads{1995ESASP.379...99B}, % Bastian
\citeads{1998RvMP...70.1393S}, % Sovers
\citeads{2003A&A...404..743K}, % Kovalevsky
\citeads{2013A&A...559A..95T}). % Titov & Lambert
This effect should produce a gradually increasing distortion
of the ICRF2 positions, in the form of a glide vector directed towards
the Galactic Centre and with an amplitude proportional to the time elapsed
since the mean epoch of the VLBI observations.
Using galactic parameters from \citetads{2009ApJ...700..137R}, and taking
the mean epoch of ICRF2 to be 13.7~years earlier than Gaia DR1, the expected
amplitude of the distortion is 0.074~mas. The observed glide is almost 
twice as large and directed ${\sim}30^\circ$ away from the Galactic Centre.
It is therefore probably a combination of the galactic effect with some 
other systematics. The unexplained (non-galactic) component is mainly 
directed along the $z$ axis. 

Conceptually, the two reference frames (Gaia in the visible and ICRF2
in the radio domain) are materialisations of the same reference
system. Our test shows that this holds globally at a level of 0.2~mas,
again a very satisfactory situation for this first Gaia solution.

\section{Discussion of the optical--radio offsets}\label{sect:discussion}

The non-coincidence of the optical and radio centres of ICRF sources has long 
been a concern for the achievement of an accurate optical reference frame
(e.g.\ \citeads{2002AJ....124..612D} and \citeads{2012MmSAI..83..952M}).
Although there are good theoretical reasons to expect that the offsets are generally small 
($<\,$1~mas; e.g.\ \citeads{2008A&A...483..759K} and \citeads{2012A&A...538A.107P}),
observational studies tend to find much larger effects (e.g.\ \citeads{2011A&A...526A..25T},
\citeads{2013A&A...553A..13O}, and the surveys discussed below). 

Based on deep CCD imaging of the optical counterparts of ICRF sources, linked to the
Hipparcos/Tycho-2 reference frame, \citetads{2014AJ....147...95Z} at USNO hypothesised 
the existence of a detrimental, astrophysical, random noise (DARN) optical--radio offset 
on a level of $\simeq\,$10~mas. Comparing Gaia DR1 optical--radio coordinate differences 
with those in Tables~4 and 5 in the USNO survey we find 
little or no correlation for the $\sim\,$320 sources in common. While the RSE of the 
coordinate differences for the common sources is less than 1~mas in Gaia DR1, it is 
20--30~mas in the USNO survey.   
A similar comparison with the Rio survey (\citeads{2013MNRAS.430.2797A}) gives the 
same result. On the other hand, there is a clear (if rather weak) positive correlation between 
the differences in the USNO and Rio surveys, although different instruments were used. 

Clearly most of the offsets found in the Rio and USNO surveys must have other causes 
than a real non-coincidence of the optical and radio centres. A reasonable hypothesis is 
that the offsets are to a large extent caused by the spatially correlated (i.e.\ systematic,
mainly zonal) errors in the Tycho-2 proper motions revealed by Gaia (see Appendix~C.1 
in \citealt{2016GaiaL}). The errors, on a level of 2--6~mas~yr$^{-1}$, are much larger 
than assumed by \citetads{2014AJ....147...95Z}, and could contribute offsets up to 
$\sim\,$50~mas over the $\sim\,$10~year epoch difference between the surveys and 
Hipparcos/Tycho-2. Although a DARN effect will surely exist at some level, it must in 
general be well below 1~mas.

\section{Conclusions}

As part of Gaia DR1 we present the optical positions of 2191 Gaia sources matched 
to ICRF2 sources, including 262 of the defining sources in ICRF2. These positions,
which come from the special auxiliary quasar solution, are shown to be more 
accurate than the positions from the secondary solution given elsewhere in Gaia DR1. 
Magnitudes in Gaia's $G$ band are given for 2152 of the sources (260 defining).
The properties of the optical data are discussed and detailed comparisons made 
between the optical and radio positions of the ICRF sources. The main conclusions are:
\begin{itemize}
\item 
The $G$ magnitudes span a range from 12.4 to 21.0~mag with the bulk of sources
between 17 and 20~mag.
\item The formal accuracy of the optical positions has a floor at $\sim\,$0.25~mas for
$G<17$~mag, gradually increasing to a few mas at $G=20$. There is no systematic difference
between defining and non-defining ICRF sources in terms of their optical accuracies versus 
magnitude.
\item  
The overall agreement between the optical and radio positions is excellent: the angular separation
is $<\,$1~mas for 44\% of the sources and $<\,$10~mas for 94\% of the sources.
For the defining sources the corresponding numbers are 71\% and 98\%.
\item
Analysis of the large-scale systematic differences between the optical and radio positions
in terms of VSH reveals no significant components except for a glide of amplitude 
$\sim\,$0.15~mas. 
\item
The angular separations are in general consistent with the combined formal uncertainties in
ICRF2 and the Gaia data, supporting the claimed accuracies. The uncertainties of the radio
positions of VCS-only sources in ICRF2 may be overestimated.
\item 
For most of the 6\% sources with angular separations above 10~mas the optical--radio
offsets are consistent with the stated formal uncertainties of the data, but for a quarter of them
the offsets are statistically significant. Individual examination of a number of these cases show 
that a likely explanation for the offset can often be found, for example in the form of a bright
host galaxy or nearby star. 
\item  
Among the sources with good optical and radio astrometry we found no indication of physical 
optical--radio offsets exceeding a few tens of mas. For most sources the true offsets are likely
to be less than 1~mas.
\end{itemize}
The last result is very encouraging for the future alignment of the very accurate optical reference 
frame to be built from Gaia observations and the corresponding radio frame, using common sources 
in two very different wavelength domains.

\begin{acknowledgements}

This work has made use of data from the ESA space mission Gaia, 
processed by the Gaia Data Processing and Analysis Consortium
(DPAC).
This research has made use of \emph{Aladin sky atlas}  and the \emph{Simbad database} developed at CDS, Strasbourg Observatory, France. We are grateful to the developers of TOPCAT (\citeads{2005ASPC..347...29T}) for their software.
Funding for the DPAC has been provided by national institutions, in
particular the institutions participating in the Gaia Multilateral Agreement.
The Gaia mission website is \url{http://www.cosmos.esa.int/gaia}.

The authors are members of the Gaia DPAC.  
This work has been supported by:

the European Space Agency in the framework of the Gaia project; 

the Centre National d'Etudes Spatiales (CNES);

the German Aerospace Agency DLR under grants 50QG0501, 50QG1401 50QG0601, 50QG0901 and 50QG1402; 

the Swedish National Space Board.

The authors are very grateful to Chris Jacobs for his constructive review and thoughtful remarks,
which helped to improve the paper.

\end{acknowledgements}
%--------------------------------------------------------------------

\bibliographystyle{aa} % style aa.bst
\bibliography{BiblioICRF} % your references Yourfile.bib
\appendix

\section{Identification of ICRF sources in the secondary solution of Gaia DR1} 
\label{sect:secondaryMatches}

Apart from the auxiliary quasar table, all positional data in Gaia DR1 for faint sources 
come from the secondary solution. For most of the sources in the auxiliary quasar table
one can find slightly different positions from the secondary solution. In this appendix we 
briefly consider the optical positions of ICRF sources in the secondary solution, even 
though they are not used in the rest of the paper.

Searching the secondary solution of Gaia DR1 for positional matches to the
3414 radio positions in ICRF2 results in 2299 matches within 150~mas (the same limit 
as was used in Sect.~\ref{sect:finalSelection}). All of the matches are unique in the sense that
there is at most one optical source within 150~mas of the radio source. Among the
ICRF sources there are 260 defining, 657 non-VCS, and 1382 VCS-only sources.

Comparing the optical positions from the two tables (auxiliary and secondary)
with the matched ICRF2 positions clearly demonstrates the superiority of the auxiliary quasar
solution. For example, the median separation (Eq.~\ref{eq:rho}) is $\rho=1.2$~mas for 
the auxiliary quasar table and 1.8~mas for the secondary table. The difference is even 
more pronounced for the defining sources, where the median separations are 
0.6~mas (auxiliary) and 1.1~mas (secondary). These results do not change significantly 
if we only consider only the subset of 2135 ICRF2 sources (of which 256 are defining) 
that appear in both tables.

Not only are the positions from the auxiliary quasar solution more accurate than in the
secondary solution, but the associated standard uncertainties also appear to be more
reliable in the auxiliary solution. This can be seen from the robust dispersion (RSE) of the normalised 
coordinate differences $X_\alpha$, $X_\delta$ (Eq.~\ref{eq:normDiff}). For the auxiliary 
quasar solution the dispersion is close to one, as expected, even for the defining 
sources (see Figs.~\ref{norm2_da_dd_histo} and \ref{norm2_da_dd_def_histo}). For
the secondary solution the dispersion is about 10\% higher for the whole sample and
30\% higher for the defining sources. This is a strong indication that the positional 
uncertainties are severely underestimated in the secondary solution, which is not the
case for the auxiliary table. There are thus good reasons to restrict the detailed
analysis of the optical--radio differences to the auxiliary quasar table.

As already mentioned, the two tables have 2135 ICRF2 sources in common, which
means that the secondary table contains $2299-2135=164$ ICRF2 sources that are
not listed in the auxiliary table. Conversely, the auxiliary table contains
$2191-2135=56$ ICRF2 sources not listed in the secondary table. The relation
between the two tables is even more complicated: among the 2135 ICRF2
sources that appear in both tables, 56 have different Gaia source identifiers in the
two tables. This means (cf.\ Sect.~\ref{subsec:gaiadetections}) that the auxiliary and
secondary positions for these 56 ICRF2 sources were computed from disjoint subsets
of the Gaia detections. Complications of this sort happened in Gaia DR1 because of the
imperfect state of the initial source list and the relatively large attitude and calibration
errors in the initial processing steps, but in later releases they will eventually be 
eliminated.

%\input{Appendix_B}
%\appendix

\section{Derivation of Eq.~(\ref{eq:normSep})} 
\label{sect:deriveqRayleigh}

\subsection{Covariance for the differences of two random vectors}
Consider two independent $k$-dimensional random vectors $\vec{x}_1$ and $\vec{x}_2$  
with mutlivariate normal distribution  with means $\vec{\mu}_1, \vec{\mu}_2 $ and covariance matrices respectively $\vec{\Sigma}_1$ and $\vec{\Sigma}_2$. Each distribution is then $\mathcal{N}_k(\vec{\mu},\vec{\Sigma})$.  Without changing the subsequent argument, one can always assume that  $\vec{\mu}_1 = \vec{\mu}_2 = 0 $.
As a consequence of the independence of the two random vectors, the joint distribution of the vector of $2k$-dimension $\vec{x} = [\vec{x}_1, \vec{x}_2]^\text{T}$ has also a multivariate normal distribution with zero mean and covariance matrix,
\begin{equation}\label{eq:covmat}
 \vec{\Sigma}^* = \begin{bmatrix}
 \vec{\Sigma}_1 & \vec{0} \\
 \vec{0} & \vec{\Sigma}_2
\end{bmatrix} \, ,
\end{equation}
where $\vec{\Sigma}^*$ is a $2k\times2k$ positive definite symmetric matrix.
Introduce now the $k$-dimensional random vector 
\begin{equation}\label{eq:difvect}
\vec{y} = \vec{x}_2 - \vec{x}_1 \, .
\end{equation}
This transformation can also be  written as linear mapping from $\mathcal{R}_{2k}$ to $\mathcal{R}_{k}$ as
\begin{equation}\label{eq:lintrans}
\vec{y} =  \vec{B} \vec{x} \, ,
\end{equation}
where $ \vec{B}$ is a $k\times 2k$ matrix, built with the $k$-dimensional unit matrix $\vec{I}_k $,
\begin{equation}\label{eq:matB}
\vec{B}=
\begin{bmatrix}
 \vec{I}_k & -\vec{I}_k 
\end{bmatrix} \, .
\end{equation}
Therefore the probability distribution of $\vec{y}$ is $\mathcal{N}_k(\vec{0},\vec{\Sigma})$ where the covariance matrix is given by
\begin{equation}
\vec{\Sigma}= \vec{B}\vec{\Sigma}^*\vec{B}^\text{T}
\end{equation}
or more trivially,
\begin{equation}\label{eq:combcov}
\vec{\Sigma}= \vec{\Sigma}_1 + \vec{\Sigma}_2 \, ,
\end{equation}
a generalisation of the addition of the variances for two scalar independent variables. 

For a normal distribution $\mathcal{N}_k(\vec{0},\vec{\Sigma})$ of a $k$-dimensional random vector $\vec{x}$  the quadratic form
\begin{equation}\label{eq:quadform}
 \vec{x}^\text{T} \vec{\Sigma}^{-1} \vec{x}
\end{equation}
follows a $\chi^2$ distribution with $k$ degrees of freedom. 
\subsection{Application to $k=2$}
This is the case of interest in this paper with the differences between Gaia and ICRF2 respectively in right ascension and declination. Let put $\vec{x}_1= (x_1, y_1) $ and $\vec{x}_2 = (x_2, y_2)$, one has successively,
\begin{equation}
\vec{\Sigma}_1 =
\begin{bmatrix}
\sigma^2_{x_1}  & c_1 \\
c_1 & \sigma^2_{y_1}
\end{bmatrix}
\end{equation}
and 
\begin{equation}
\vec{\Sigma}_2 =
\begin{bmatrix}
\sigma^2_{x_2}  & c_2 \\
c_2 & \sigma^2_{y_2}
\end{bmatrix}
\end{equation}
and finally,
\begin{equation}
\vec{\Sigma} =
\begin{bmatrix}
\sigma^2_{x_1} +\sigma^2_{x_2}  & c_1 +c_2 \\
c_1 +c_2 & \sigma^2_{y_1}+\sigma^2_{y_2}
\end{bmatrix}
\end{equation}
where $c_1 = \text{cov}(x_1, y_1)$ and $c_2 = \text{cov}(x_2, y_2)$, respectively the covariance between the errors in the ICRF2 and in the Gaia solution in right ascension and declination and then $x_\alpha = x_2- x_1$ and $x_\delta = y_2- y_1$ for the two components of the positional differences.

The quadratic form
\begin{equation}\label{eq:formQ}
 Q_{\alpha, \delta} = \begin{bmatrix} x_\alpha & x_\delta\end{bmatrix} \vec{\Sigma}^{-1} \begin{bmatrix}  x_\alpha \\ x_\delta\end{bmatrix}
\end{equation}
therefore follows a $\chi^2$ distribution with 2 degrees of freedom, or equivalently $\sqrt{Q_{\alpha, \delta}}$ follows a standard Rayleigh distribution. An analytical inversion of $\vec{\Sigma}$ gives the same as 
Eq.~(\ref{eq:normSep}), although it is written with reduced variables with unit variances and the covariance identical to the correlation coefficient. Thus $C$ in Eq.~(\ref{eq:normC}) is the same as
\begin{equation}\label{eq:fullC}
 \frac{c_1+c_2}{\sqrt{\sigma^2_{x_1} +\sigma^2_{x_2}}\sqrt{\sigma^2_{y_1} +\sigma^2_{y_2}}} \, .
\end{equation}

\listofobjects

\end{document}